\PassOptionsToPackage{letterpaper,margin=0cm,bottom=0.1cm}{geometry}
\documentclass[prx,11pt,letterpaper,twocolumn,unpublished]{quantumarticle}
\pdfoutput=1
\usepackage[letterpaper,margin=0cm,bottom=0.1cm]{geometry}

\usepackage{xcolor}
\definecolor{darkblue}{RGB}{40, 85, 120} 
\definecolor{quantumviolet}{HTML}{53257F} 
\definecolor{quantumgray}{HTML}{555555} 
\definecolor{quantumgreen}{HTML}{007474} 
\definecolor{quantumblue}{HTML}{3A5FCD} 
\definecolor{quantumpurple}{HTML}{8A2BE2} 
\definecolor{warmorange}{HTML}{FF8C42} 
\definecolor{violetpurple}{HTML}{8A2BE2} 
\definecolor{teal}{HTML}{008080} 
\definecolor{softbeige}{HTML}{ECE5D7} 
\definecolor{charcoalgray}{HTML}{2E2E2E} 
\definecolor{coralred}{HTML}{FF6F61} 
\definecolor{brightyellow}{HTML}{FFD700} 
\definecolor{cyanblue}{HTML}{1C77C3} 
\definecolor{deepbluegray}{HTML}{2C3E99} 
\definecolor{softblue}{HTML}{AFCBFF} 

\usepackage{amsmath}
\usepackage{amssymb}
\usepackage{amsthm}
\usepackage{bm}
\usepackage{braket}
\usepackage{xcolor}
\usepackage{graphicx}
\usepackage[caption=false]{subfig}
\usepackage[colorlinks, linktocpage, hypertexnames = false]{hyperref}

\hypersetup{linkcolor = {gray!80!black},
citecolor = {blue!90!black}, 
urlcolor = {gray!60!black}}

\newcommand{\orcidd}[1]{\href{https://orcid.org/#1}{\includegraphics[width=8pt]{orcid}}} 

\usepackage{tikz}
\usetikzlibrary{positioning,intersections}
\usetikzlibrary{calc}
\usetikzlibrary{shapes.geometric}
\usepackage{braids}
\usepackage{tikz-cd}
\usetikzlibrary{shapes.geometric,shapes.misc}
\usetikzlibrary{arrows,matrix,calc,scopes,decorations.markings,snakes}
\usetikzlibrary{arrows.meta}
\usetikzlibrary{intersections, patterns,fit} 
\usetikzlibrary{decorations.pathreplacing,angles,quotes}
\usetikzlibrary{decorations.markings}

\tikzset{
    partial ellipse/.style args={#1:#2:#3}{
        insert path={+ (#1:#3) arc (#1:#2:#3)}
    }
}

\tikzset{->-/.style={decoration={
  markings,
  mark=at position #1 with {\arrow{>}}},postaction={decorate}}}
\tikzset{-<-/.style={decoration={
  markings,
  mark=at position #1 with {\arrow{<}}},postaction={decorate}}}

\usetikzlibrary{decorations.markings}

\theoremstyle{theorem}

\theoremstyle{remark}

\newcommand{\Av}{\mathbb{A}}
\newcommand{\Bf}{\mathbb{B}}

\newcommand{\Hbb}{\mathbb{H}}

\newcommand{\Tsing}{\mathcal{T}_{\mathsf{Ising}}}
\newcommand{\Tube}{\mathbf{Tube}}

\newcommand{\one}{\mathbb{1}}

\newcommand{\Rep}{\mathsf{Rep}}

\newcommand{\XR}{\overset{\rightarrow}{X}}
\newcommand{\XL}{\overset{\leftarrow}{X}}

\newcommand{\ZR}{\overset{\rightarrow}{Z}}
\newcommand{\ZL}{\overset{\leftarrow}{Z}}

\newcommand{\Cocom}{\operatorname{Cocom}}

\newcommand{\cone}{(1)}
\newcommand{\ctwo}{(2)}
\newcommand{\cthree}{(3)}

\newcommand{\id}{\operatorname{id}}
\newcommand{\Irr}{\operatorname{Irr}}
\newcommand{\Hom}{\operatorname{Hom}}

\newcommand{\Fun}{\mathsf{Fun}}
\newcommand{\Vect}{\mathsf{Vect}}
\newcommand{\Ising}{\mathsf{Ising}}
\newcommand{\TIsing}{\mathcal{T}_{\mathsf{Ising}}}
\newcommand{\hTIsing}{\hat{\mathcal{T}}_{\mathsf{Ising}}}

\newcommand{\cA}{\text{\usefont{OMS}{cmsy}{m}{n}A}}

\newcommand{\cS}{\text{\usefont{OMS}{cmsy}{m}{n}S}}

\usepackage{euscript}

\newcommand\eC           {\EuScript{C}}

\newcommand\eM          {\EuScript{M}}

\newcommand\eS         {\EuScript{S}}

\usepackage[bbgreekl]{mathbbol} 
\DeclareMathAlphabet{\mathcal}{OMS}{cmsy}{m}{n}

\usepackage[noframe]{showframe}
\usepackage{framed}
\usepackage{lipsum}
 {\endMakeFramed}
\definecolor{shadecolor}{gray}{0.9} 

\usepackage{multirow}

\usepackage{array}
\newcolumntype{R}[1]{>{\raggedright\arraybackslash}p{#1}}

\usepackage{orcidlink} 

\begin{document}

\flushbottom

\title{Quantum Cluster State Spin Chain with Ising Fusion Category Symmetry: A Perspective from Weak Hopf SymTFT}

\author{Zhian Jia\orcidlink{0000-0001-8588-173X}}
\email{giannjia@foxmail.com}
\affiliation{Centre for Quantum Technologies, National University of Singapore, SG 117543, Singapore}
\affiliation{Department of Physics, National University of Singapore, SG 117543, Singapore}


\begin{abstract}
In this work, we present a construction of a cluster state lattice Hamiltonian that exhibits the symmetry of the Ising fusion algebra. This construction is formulated within the framework of weak Hopf symmetry topological field theory (SymTFT), where we assign smooth and rough boundaries to the weak Hopf quantum double model, thereby extending the conventional cluster state model.
Central to our construction is the weak Hopf Ising boundary tube algebra $\mathcal{T}_{\mathsf{Ising}}$, whose representation category is equivalent to the Ising fusion category $\mathsf{Ising}$. We take this algebra as the input data for the weak Hopf quantum double model.
The resulting model exhibits Ising fusion symmetry on both open and closed $1\text{d}$ manifolds. On open manifolds, the symmetry is governed by $\mathcal{T}_{\mathsf{Ising}} \otimes \mathcal{T}_{\mathsf{Ising}}^{\vee}$; on closed manifolds, it reduces to $\operatorname{Cocom}(\mathcal{T}_{\mathsf{Ising}}) \otimes \operatorname{Cocom}(\mathcal{T}_{\mathsf{Ising}}^{\vee})$. Since the Ising fusion algebra embeds into $\operatorname{Cocom}(\mathcal{T}_{\mathsf{Ising}}^{\vee})$, the model faithfully realizes the symmetry of the Ising fusion category.
\end{abstract}

\maketitle

\tableofcontents

\section{Introduction}
\label{sec:intro}

Symmetry is a central concept in physics. Traditionally, it is characterized by group structures; however, in recent years, this notion has been generalized in various directions (see, e.g.,~\cite{cordova2022snowmass,brennan2023introduction,mcgreevy2023generalized,luo2023lecture,gomes2023introduction,shao2024whats,SchaferNameki2024ICTP,Bhardwaj2024lecture}). A key insight is that symmetry can be understood in terms of topological defects~\cite{gaiotto2015generalized}. In this framework, symmetries are realized either via \emph{symmetry operators}, which are supported on spatial submanifolds, or \emph{symmetry defects}, whose support includes a temporal direction.

Three key ingredients of generalized symmetry, as emphasized in~\cite{Cao2023subsystem}, are: the codimension of the manifold supporting the symmetry; the (non-)invertibility and unitarity of the symmetry action; and the topological nature of the symmetry. Depending on how these features are specified, several notable generalizations of symmetry have been developed.
The manifold supporting a group symmetry can be extended to higher codimensions, giving rise to higher-form symmetries~\cite{gaiotto2015generalized,kapustin2017higher,gomes2023introduction,Bhardwaj2024lecture}. The group structure itself can be extended to higher groups, which remain invertible~\cite{Cordova2019highergroup}. 
More generally, the algebraic structure underlying symmetry can be promoted to fusion categories or higher fusion categories, leading to \emph{non-invertible symmetries} (see~\cite{cordova2022snowmass,brennan2023introduction,mcgreevy2023generalized,luo2023lecture,shao2024whats,SchaferNameki2024ICTP,Bhardwaj2024lecture} and references therein). Further extensions include Hopf algebras, weak Hopf algebras, and other quantum algebras that capture the diverse symmetry structures realized in quantum systems~\cite{Buerschaper2013a,meusburger2017kitaev,chen2021ribbon,jia2023boundary,Jia2023weak,jia2024generalized,jia2024weakTube,Choi2024duality,jia2024weakhopfnoninvertible,jia2024quantumcluster,meng2024noninvertiblespt,inamura2022lattice,inamura2023fermionization}.
While these generalizations typically preserve the requirement that symmetry operators be topological, this condition can also be relaxed and one obtains the notion of \emph{subsystem symmetry}, \emph{modulated symmetry}, etc. (see, e.g.,~\cite{You2018SSPT,Devakul2018classifcation,Shirley2019foliated,jia2025subsystemsymmetry,Cao2023subsystem} and references therein).

\begin{figure}[b]
    \centering
    \includegraphics[width=1\linewidth]{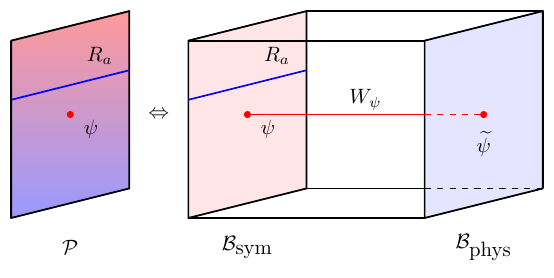}
\caption{Illustration of SymTFT, where the symmetry and dynamics of a $(d+1)$D theory are encoded on separate boundaries of a $(d+2)$D topological field theory.}
    \label{fig:SymTFT}
\end{figure}

For an $n$-dimensional quantum field theory, non-invertible symmetry is characterized by a fusion $n$-category $\mathcal{S}$. From the perspective of topological holography or the SymTFT framework~\cite{Kong2020algebraic,luo2023lecture,gomes2023introduction,shao2024whats,SchaferNameki2024ICTP,Bhardwaj2024lecture,huang2023topologicalholo,bhardwaj2024lattice,freed2024topSymTFT,gaiotto2021orbifold,bhardwaj2023generalizedcharge,apruzzi2023symmetry,bhardwaj2024gappedphases,Zhang2024anomaly,Ji2020categoricalsym,bhardwaj2024clubsandwich,bhardwaj2024hassediagramsgaplessspt}, the fusion $n$-symmetry $\mathcal{S}$ corresponds to a topological boundary condition of an $(n+1)$-dimensional topological order, described by the Drinfeld center $\mathcal{Z}(\mathcal{S})$.  See Figure~\ref{fig:SymTFT} for an illustration.
In general, the topological order may also admit a second boundary, referred to as the physical boundary $\mathcal{B}_{\text{phys}}$, which encodes the non-topological (i.e., physical) degrees of freedom of the system. This boundary $\mathcal{B}_{\text{phys}}$ can be either gapped or gapless. Upon dimensional reduction of the $(n+1)$-dimensional bulk theory with boundaries, one recovers the original $n$-dimensional theory with symmetry $\mathcal{S}$.

In the special case $n = 1$, the categorical symmetry is described by a fusion category $\mathcal{S}$ (i.e., a fusion 1-category). The symmetry is said to be \emph{anomaly-free} if there exists a one-dimensional $\mathcal{S}$-SPT phase; otherwise, it is referred to as \emph{anomalous}. For an anomaly-free fusion category $\mathcal{S}$, the associated $\mathcal{S}$-SPT phases are classified by fiber functors $F: \mathcal{S} \to \mathsf{Vect}$, i.e., tensor functors from $\mathcal{S}$ to the category of finite-dimensional vector spaces~\cite{thorngren2019fusion}.

A crucial and challenging question, which is currently under intensive investigation, is whether it is possible to construct a lattice model that realizes a given $\mathcal{S}$-symmetric phase~\cite{aasen2016topological,aasen2020topological,inamura2022lattice,Inamura2024fusionSuface,eck2024generalizationskitaevshoneycombmodel,fechisin2023noninvertible,bhardwaj2024lattice,bhardwaj2024gappedphases,Feiguin2007interacting,trebst2008short,buican2017anyonic,Lootens2023dualityHamiltonian,Lootens2024duality,Seiberg2024noninvertible,Seifnashri2024cluster,inamura202411dsptphasesfusion,cordova2024noninvertiblesymmetriesfinitegroup}.
It is important to note that the notion of fusion category symmetry in lattice models is more subtle than in continuum quantum field theories~\cite{Seiberg2024noninvertible}. In particular, one must carefully distinguish between \emph{symmetry operators} and \emph{symmetry defects}. A symmetry operator acts on the Hilbert space of the theory, typically implementing a global transformation, whereas a symmetry defect manifests as a local modification of the Hamiltonian, often altering the couplings along a codimension-one submanifold. In this work, we focus primarily on symmetry operators.
Another point worth emphasizing is that, in the context of lattice models, attention is often directed toward the fusion algebra $K_0(\mathcal{S})$—also known as the Grothendieck ring—associated with the fusion category $\mathcal{S}$. This algebra is generated by the isomorphism classes of simple objects in $\mathcal{S}$, with multiplication defined by the fusion rules:
\begin{equation}
    [a] \cdot [b] := [a \otimes b] = \sum_{c \in \Irr(\mathcal{S})} N_{ab}^c\, [c],
\end{equation}
where $N_{ab}^c$ are the fusion coefficients and $\Irr(\mathcal{S})$ denotes the set of simple objects in $\mathcal{S}$.
A Hamiltonian $H:\mathcal{H}\to \mathcal{H}$ is said to have $\cS$ symmetry if there is a representation $R: K_0(\cS)\to \operatorname{End}(\mathcal{H})$ supported by Hilbert space  $\mathcal{H}$  and $[R_{[a]},H]=0$ for all $[a]\in K_0(\mathcal{S})$.
This algebraic description of non-invertible symmetry via the fusion algebra erases certain information, particularly that related to anomalies. For example, the categories $\Vect_G$ and $\Vect_G^{\omega}$ share the same fusion algebra, yet they differ at the level of symmetry: the latter corresponds to a $G$-symmetry with a nontrivial 't Hooft anomaly classified by $\omega \in H^3(G, U(1))$. The key distinction lies in the associator data of the fusion category, which captures the anomaly but is invisible if one only considers the fusion algebra $K_0(\mathcal{S})$. In this sense, the fusion algebra provides a coarse invariant that may obscure important categorical information such as anomalies. A complete description of fusion category symmetry should also include symmetry defects and their interplay with symmetry operators~\cite{Seiberg2024noninvertible}. Through this extended structure, additional information—such as anomalies—can be encoded and detected.
Nevertheless, studying the fusion algebra symmetry remains important, as it serves as a natural starting point for understanding non-invertible symmetries on the lattice.

In our previous works, we developed the weak Hopf SymTFT framework as a systematic approach to constructing lattice models with a prescribed non-invertible symmetry~\cite{jia2024generalized,jia2024weakhopfnoninvertible}. A notable feature of this framework is that the Hilbert space is always a tensor-product space. Moreover, since the construction focuses only on the fusion algebra of a given fusion category, anomalous symmetries—such as those associated with the Haagerup fusion category—can also be realized within this class of lattice models~\cite{jia2024quantumcluster}. As illustrated above, this does not lead to any contradiction, since the information of anomaly is erased at the fusion algebra level.
Another point we would like to emphasize is that, within this framework, the non-invertible symmetry is typically not on-site\footnote{The notion of ``on-siteness'' (or locality) for non-invertible symmetries requires further clarification; see~\cite{meng2024noninvertiblespt}. To our knowledge, this problem remains open.} in the usual sense, yet it maps local operators of the Hamiltonian to local operators.

In this short note, we take the non-invertible symmetry characterized by the Ising fusion category $\Ising$ as a concrete example to further illustrate our method. Since there exists no fiber functor from $\Ising$ to $\mathsf{Vect}$, the category $\Ising$ is regarded as an anomalous symmetry and is therefore generally believed to be difficult to realize on a tensor-product Hilbert space.
Our construction primarily relies on the weak Hopf tube algebra developed in our recent work~\cite{jia2025weakhopftubealgebra}, which enables the reconstruction of a weak Hopf algebra whose representation category is equivalent to $\Ising$. We then take this weak Hopf algebra as the input data for a weak Hopf quantum double model, and impose a set of topological boundary conditions that encode the $\Ising$ symmetry as a smooth boundary. In this way, we realize a lattice model that faithfully implements the $\Ising$ symmetry.
As pointed out in Ref.~\cite{jia2024generalized}, the standard cluster state model can be interpreted as a quantum double model with one smooth boundary and one rough boundary. Following this perspective, we construct the Ising cluster state model in an analogous manner.

This note is organized as follows.
In Section~\ref{sec:IsingSym}, we introduce a weak Hopf algebra realization of the $\Ising$ symmetry, constructed from the Ising boundary tube algebra. We discuss how this algebra captures the symmetry, and how its dual and associated quantum double can be understood through the tube algebra formalism.
In Section~\ref{sec:IsingCluHam}, we construct the Ising cluster state model and demonstrate that it exhibits symmetry governed by the fusion category $\Ising$.
Section~\ref{sec:IsingCluStateTN} presents an exact solution of the model using the weak Hopf tensor network formalism, identifying the ground state as the Ising cluster state.
In the final section, we offer some concluding remarks.

\section{Ising Fusion Category Symmetry and Its Weak Hopf Realization}
\label{sec:IsingSym}

We begin by reviewing the general properties of the Ising fusion category symmetry and explaining how it can be understood in terms of a weak Hopf symmetry. The presentation in this section is intended to be as self-contained as possible. For further details on related concepts and the conventions adopted here, we refer the reader to \cite{Jia2023weak,jia2024generalized,jia2024weakTube,jia2025weakhopftubealgebra}.

\subsection{Ising fusion category symmetry}

The Ising fusion category $\Ising$ originates from the Ising conformal field theory; for an introductory explanation, see, e.g.,~\cite[Section II.7]{ketov1995conformal}.
It contains three simple objects: $\mathbb{1}$, $\sigma$, and $\psi$, all of which are self-dual. The non-trivial fusion rules are given by (see Table~\ref{tab:IsingFusion} for the fusion table)
\begin{equation}
    \sigma \otimes \psi = \psi \otimes \sigma = \sigma, \quad
    \psi \otimes \psi = \mathbb{1}, \quad
    \sigma \otimes \sigma = \mathbb{1} \oplus \psi.
\end{equation}
The objects $\mathbb{1}$ and $\psi$ are abelian, thus their quantum dimensions are $d_{\mathbb{1}} = d_{\psi} = 1$. From the fusion rule $\sigma \otimes \sigma = \mathbb{1} \oplus \psi$, it follows that the quantum dimension of $\sigma$ is $d_{\sigma} = \sqrt{2}$.
Moreover, the category $\mathsf{Ising}$ admits a braiding structure and is, in fact, a modular tensor category \cite{bakalov2001lectures}. In particular, this implies that fusion is commutative, which is already evident from the explicit form of the fusion rules.

\begin{table}[t]
\caption{\label{tab:IsingFusion}Fusion rules for the Ising fusion category. Elements of the first column are fused with elements of the first row (``column labels'' $\otimes$ ``row labels'').}
\centering
\begin{tabular}{l|c|c|c}
\hline
\hline
$\otimes$ & $\mathbb{1}$ & $\psi$ & $\sigma$ \\
\hline
$\mathbb{1}$ & $\mathbb{1}$ & $\psi$ & $\sigma$ \\
\hline
$\psi$ & $\psi$ & $\mathbb{1}$ & $\sigma$ \\
\hline
$\sigma$ & $\sigma$ & $\sigma$ & $\mathbb{1} \oplus \psi$ \\
\hline
\hline
\end{tabular}
\end{table}

Since fusion multiplicity is one, namely $\dim \Hom(a\otimes b,c)=1$ for all $a,b,c\in \Irr(\Ising)$, there is no need to label the fusion/splitting vertex.
The fusion/splitting vertex can be represented as
\begin{equation}
    \begin{aligned}
        %
 
    \end{aligned} \in \Hom(c,a\otimes b).
\end{equation}
The three basic topological local moves are used \cite{jia2024weakTube,jia2025weakhopftubealgebra} in order to construct the boundary tube algebra. First, the nontrivial loop moves (those that do not involve $\one$) are 
\begin{equation}
    \begin{aligned}
        %
    \end{aligned}\;\;.
\end{equation}
Second, the nontrivial parallel moves are 
\begin{align}
   & \begin{aligned}
        %
    \end{aligned}. 
\end{align}
The F-moves are also easy to written down. Three $\sigma$ can only fuse to $\sigma$, we have the following 
\begin{align}
    &\begin{aligned}
        %
    \end{aligned}. 
\end{align}
All other $F$-symbols are $\pm 1$, with two nontrivial sign for fusion tree of $\psi \to \sigma \otimes \psi \otimes \sigma$ and $\sigma \to \psi \otimes \sigma \otimes \psi$ 
\begin{align}
    &\begin{aligned}
        %
    \end{aligned}.
\end{align}

The Ising fusion ring (i.e., the Grothendieck ring of $\Ising$), denoted by $K_0(\Ising)$, is a $\mathbb{Z}_{\geq 0}$-ring with a basis given by the simple objects in $\Ising$, and multiplication defined via the fusion rules.
A representation $Q_{\one}, Q_{\sigma}, Q_{\psi} \in \operatorname{End}(\mathcal{H})$ on a Hilbert space $\mathcal{H}$ satisfies the fusion relations:
\begin{equation}
    Q_{\one} = I,  Q_{\sigma} Q_{\psi} = Q_{\sigma} = Q_{\psi} Q_{\sigma},  Q_{\psi}^2 = I, Q_{\sigma}^2 = I + Q_{\psi}.
\end{equation}
A Hamiltonian $H$ on $\mathcal{H}$ is said to have $\Ising$ symmetry if and only if
\begin{equation}
    [H, Q_{\psi}] = [H, Q_{\sigma}] = 0.
\end{equation}

For instance, consider a one-dimensional representation given by $Q_{\mathbb{1}} = 1$, $Q_{\psi} = -1$, and $Q_{\sigma} = 0$. In this case, any real-valued Hamiltonian is symmetric under the $\Ising$ fusion algebra. Another example is the representation $Q_{\mathbb{1}} = 1$, $Q_{\psi} = 1$, $Q_{\sigma} = \sqrt{2}$.
On a qubit Hilbert space, we can realize the representation as
\[
Q_{\mathbb{1}} =  I, 
Q_{\psi} = \sigma_x = \begin{pmatrix}
0 & 1 \\
1 & 0
\end{pmatrix}, 
Q_{\sigma} = \frac{1}{\sqrt{2}} \begin{pmatrix}
1 & 1 \\
1 & 1
\end{pmatrix}.
\]
A Hamiltonian of the form
\[
H = a I + r \sigma_x = \begin{pmatrix}
a & r \\
r & a
\end{pmatrix}, \quad a, r \in \mathbb{R}
\]
commutes with both $Q_{\psi}$ and $Q_{\sigma}$, and thus exhibits $\Ising$ symmetry.
On the three-dimensional Hilbert space $\mathcal{H} = \mathbb{C}^3$, we may consider the representation
\[
Q_{\mathbb{1}} =I, 
Q_{\psi} =
\begin{pmatrix}
0 & 1 & 0 \\
1 & 0 & 0 \\
0 & 0 & 1
\end{pmatrix},
Q_{\sigma} =
\begin{pmatrix}
0 & 0 & 1 \\
0 & 0 & 1 \\
1 & 1 & 0
\end{pmatrix}.
\]
Then any Hamiltonian of the form
\[
H =
\begin{pmatrix}
a & u & v \\
u & a & v \\
v & v & a + u
\end{pmatrix}, \quad a, u, v \in \mathbb{R}
\]
commutes with both $Q_{\psi}$ and $Q_{\sigma}$, and hence realizes the $\Ising$ fusion category symmetry.

\subsection{Ising boundary tube algebra}

The main tool we use is the notion of \emph{weak Hopf symmetry}~\cite{Jia2023weak,jia2024generalized,jia2024weakTube,jia2025weakhopftubealgebra}. By definition, a complex \emph{weak Hopf algebra}~\cite{bohm1996coassociative} is a complex vector space $\mathcal{A}$ equipped with the following structure maps:
\begin{itemize}
    \item multiplication $\mu: \mathcal{A} \otimes \mathcal{A} \to \mathcal{A}$,
    \item unit $\eta: \mathbb{C} \to \mathcal{A}$,
    \item comultiplication $\Delta: \mathcal{A} \to \mathcal{A} \otimes \mathcal{A}$,
    \item counit $\varepsilon: \mathcal{A} \to \mathbb{C}$,
    \item antipode $S: \mathcal{A} \to \mathcal{A}$,
\end{itemize}
satisfying a set of compatibility conditions. We assume that $\{e_k\}_{k=0}^{d-1}$ is a basis of $\mathcal{A}$, and for convenience in many cases, we can fix $e_0 = 1$, the identity element of $\mathcal{A}$.

The axioms of a weak Hopf algebra can be summarized as follows:
\begin{enumerate}
    \item \textit{Algebra axiom:} $(\mathcal{A}, \mu, \eta)$ is an associative unital algebra, i.e.,
    \begin{align} \label{eq:WeakAlgeAxiom}
        \mu \circ (\mu \otimes \id) = \mu \circ (\id \otimes \mu),\\ 
        \mu \circ (\eta \otimes \id) = \id = \mu \circ (\id \otimes \eta).
    \end{align}
    
    \item \textit{Coalgebra axiom:} $(\mathcal{A}, \Delta, \varepsilon)$ is a coassociative counital coalgebra, but with weaker unit constraints:
    \begin{align} \label{eq:WeakCoAlgeAxiom}
        (\Delta \otimes \id) \circ \Delta = (\id \otimes \Delta) \circ \Delta, \\
        (\varepsilon \otimes \id) \circ \Delta = \id = (\id \otimes \varepsilon) \circ \Delta.
    \end{align}

    \item \textit{Weak bialgebra axiom:} The maps $\Delta$ and $\varepsilon$ are compatible with the multiplication in the following weaker sense:
    \begin{equation} \label{eq:WeakBialgAxiom}
        \Delta(ab) = \Delta(a) \Delta(b), \quad \forall a,b \in \mathcal{A}, \\
    \end{equation}
   \begin{equation}
        \begin{aligned}
            & (\Delta(1)\otimes 1)(1\otimes \Delta(1)) \\ 
             =&(\Delta\otimes \id)\circ \Delta(1) \\
             = &(1\otimes \Delta(1)) (\Delta(1)\otimes 1),
        \end{aligned}
    \end{equation} 
   \begin{align}     
        \begin{aligned}
              \varepsilon(abc) 
              = & \varepsilon(ab^{(1)})\, \varepsilon(b^{(2)}c) \\
              =& \varepsilon(ab^{(2)})\, \varepsilon(b^{(1)}c), \quad \forall a,b,c \in \mathcal{A},
        \end{aligned}
    \end{align}
    using Sweedler's notation $\Delta(b) = \sum b^{(1)} \otimes b^{(2)}$.

    \item \textit{Antipode axiom (weakened):} The antipode $S$ satisfies the identities
    \begin{align}
     &  \mu \circ (S \otimes \id) \circ \Delta = \varepsilon_R, \\ &\mu \circ (\id \otimes S) \circ \Delta = \varepsilon_L,\\
    & S  = \mu_2 \circ (S \otimes \mathrm{id} \otimes S) \circ  \Delta_2,      \label{eq:WeakAntipodeAxiom}
  \end{align}
    where the maps $\varepsilon_L, \varepsilon_R: \mathcal{A} \to \mathcal{A}$ are the canonical left and right counit maps, given by
    \begin{align}
        \varepsilon_L(x) &= (\id \otimes \varepsilon)\left[ \Delta(1) \cdot (x \otimes 1) \right], \\
        \varepsilon_R(x) &= (\varepsilon \otimes \id)\left[ (1 \otimes x) \cdot \Delta(1) \right].
    \end{align}
  We use the shorthand notation \(\mu_2 = \mu \circ (\mu \otimes \mathrm{id})\) for iterated multiplication and \(\Delta_2 = (\Delta \otimes \mathrm{id}) \circ \Delta\) for iterated comultiplication. The associativity of \(\mu\) and coassociativity of \(\Delta\) ensure that these expressions are unambiguous.
\end{enumerate}

Since any fusion category can be realized as the representation category of a weak Hopf algebra \cite{szlachanyi2000finite,ostrik2003module}, the Ising fusion category $\mathsf{Ising}$ admits such a realization as well. That is, there exists a weak Hopf algebra $\cA_{\mathrm{Ising}}$ such that $\Rep(\cA_{\mathrm{Ising}}) \simeq_{\otimes} \mathsf{Ising}$. We refer to $\cA_{\mathrm{Ising}}$ as an \emph{Ising weak Hopf algebra}. 
It is important to note that this realization is not unique: there exist many weak Hopf algebras whose representation categories are monoidal equivalent to $\mathsf{Ising}$, and these algebras are all Morita equivalent.

There are several approaches to constructing such a weak Hopf algebra. The standard method is via \emph{Tannaka–Krein reconstruction}. However, this approach generally does not yield an explicit description of the basis elements and structure constants of the algebra. In this work, we adopt a different method based on the notion of the \emph{boundary tube algebra}~\cite{Kitaev2012boundary,bridgeman2023invertible,jia2025weakhopftubealgebra,jia2024weakTube,bai2025weakhopf}.

Consider a UFC  $\EuScript{C}$ that characterize the $2d$ bulk (input data for string-net model), the gapped boundary described by a $ \EuScript{C}$-module category ${_{\EuScript{C}}}\EuScript{M}$. 
We construct the corresponding boundary tube algebra \( \mathbf{Tube}({_{\EuScript{C}}}\EuScript{M}) \), which is a \( C^* \) weak Hopf algebra. The basis of the algebra is given by
\begin{equation}
\left\{  
\begin{aligned}

\end{aligned}
:\,  \begin{aligned}
    &a\in \Irr(\eC),c,e,f,g\in \Irr(\eM),\\
    &\mu \in \Hom(a\otimes f,g),\\
    &\nu\in \Hom(c,a\otimes e) 
\end{aligned}
\right\}.
\label{eq:tubebasis}
\end{equation}
Hereinafter, we assume that all strings are oriented upwards, and we omit all arrows whenever there is no ambiguity.
We will call $a$ bulk string and $c,e,f,g$ boundary string.
The unit is give by
\begin{equation}\label{eq:WHA1}
1=\sum_{f,e\in \Irr(\eC)}
\begin{aligned}
    %
\end{aligned}.
\end{equation}
which means we set $a=\mathbb{1}$ and $c=e$, $f=g$ and $\mu=\id$, $\nu=\id$ in Eq.~\eqref{eq:tubebasis}.
The multiplication is defined as 
\begin{equation}
    \begin{aligned}
    %
\end{aligned}.
\end{equation}
After evaluating the diagram on the right-hand side using \( F \)-moves, parallel moves, and loop moves \cite{jia2025weakhopftubealgebra}, we obtain a linear combination of the basis elements in Eq.~\eqref{eq:tubebasis}.
The counit is defined by
\begin{equation}
    \varepsilon\left( 
    \begin{aligned}
    %
\end{aligned} 
.
\end{equation}
After evaluation of right-hand side, we obtain $\delta_{f,e} \delta_{c,g} \sqrt{\frac{d_a d_e}{d_g}}$.
The comultiplication is defined as
\begin{equation}
 \begin{aligned}
   &   \Delta\left( 
    \begin{aligned}
    %
\end{aligned}.
\end{aligned}
\end{equation}
The antipode map is defined as
\begin{equation}
         S\left( 
    \begin{aligned}
    %
\end{aligned}.
\end{equation}
The boundary tube algebra $\mathbf{Tube}({}_{\eC}\eM)$ is also equipped with a $*$-operation
\begin{equation} \label{eq:WHA6}
        \left( 
    \begin{aligned}
    %
\end{aligned}.
\end{equation}

If we set $\eC=\Ising$ and $\eM=\Ising$, we obtain the \emph{Ising boundary tube algebra} denoting $\mathcal{T}_{\Ising}:=\mathbf{Tube}({}_{\Ising}\Ising)$, which means we are considering the smooth boundary of Ising string-net model \cite{jia2025weakhopftubealgebra}. 
The basis can be write down explicitly as follows (grouped via bulk string type):
\begin{align*}
    &\begin{aligned}
    %
    \end{aligned} \;\; 
\end{align*}
We see that the dimension of $\mathcal{T}_{\Ising}$ is $34$. The structure constant for $\mathcal{T}_{\Ising}$ can be calculated explicitly via Eq.~\eqref{eq:WHA1}-\eqref{eq:WHA6}.

\subsection{Dual Ising boundary tube algebra}

Notice the smooth boundary excitation characterized by the category $\eS=\Fun_{\Ising}(\Ising,\Ising)$ of functor of module category $_{\Ising}\Ising$, this is equivalent to $\Ising$, namely $\Fun_{\Ising}(\Ising,\Ising) \simeq \Ising$ as fusion categories.
Using $\mathcal{T}_{\Ising}$, we can identity $\eS$ with the representation category (as fusion categories)
\begin{equation}
    \Ising \simeq_{\otimes} \Fun_{\Ising}(\Ising,\Ising)  \simeq_{\otimes} \Rep(\mathcal{T}_{\Ising}).
\end{equation}
This is generally believed to be true; see, for instance, a recently announced proof in~\cite{bai2025weakhopf}. We also establish a monoidal embedding of a given fusion category into the representation category of a weak Hopf algebra associated with the domain wall tube algebra in~\cite{jia2025weakhopftubealgebra}, using the string diagram calculus. A similar argument applies to the boundary case as well.

For all simple objects $\one,\psi,\sigma$ there is an irreducible representation $\rho_{\one},\rho_{\psi},\rho_{\sigma}$ in $\Rep(\mathcal{T}_{\Ising})$.
Recall that a representation of weak Hopf algebra $\mathcal{T}_{\Ising}$ is a pair $(\rho,\mathcal{V})$ with $\rho:\mathcal{T}_{\Ising} \to \mathrm{End}(\mathcal{V})$ being an algebraic map, meaning:
\begin{equation}
    \rho(gh)=\rho(g)\rho(h), \quad \rho(1)=1.
\end{equation}
The fusion of two representations $(\rho_1,\mathcal{V}_1)$ and $(\rho_2,\mathcal{V}_2)$ is defined via coalgebra structure of $\mathcal{T}_{\Ising}$:
\begin{equation}
    h \triangleright (v_1\otimes v_2) := \sum_{(h)} (h^{\cone} \triangleright v_1) \otimes (h^{\ctwo} v_2).
\end{equation}
and 
\begin{equation}
    \mathcal{V}_1\otimes_{\mathcal{T}_{\Ising}}\mathcal{V}_2:= \{ \Delta(1) x=x|x\in   \mathcal{V}_1\otimes_{\mathbb{C}}\mathcal{V}_2\}.
\end{equation}
The trivial representation of $\mathcal{T}_{\Ising}$ is defined as $\varepsilon_L(\mathcal{T}_{\Ising})$ where $\varepsilon_L(h)=\sum_{(1)}\varepsilon(1^{\cone}h)1^{\ctwo}$.
For readers not familiar with quantum groups, we recall that the Sweedler notation for comultiplication $\sum_{(h)}h^{\cone}\otimes h^{\ctwo}$ means $\sum_i h^{\cone}_i \otimes h_i^{\ctwo} \in \mathcal{T}_{\Ising}\otimes \mathcal{T}_{\Ising}$, for more details about convention and notation we use, see \cite{jia2023boundary,Jia2023weak,jia2024generalized,jia2024weakhopfnoninvertible}.

The dual Ising boundary tube algebra $\hat{\mathcal{T}}_{\Ising}$ consists of functionals over $\mathcal{T}_{\Ising}$:
\begin{equation}
  \hat{\mathcal{T}}_{\Ising}=\{\varphi:   \mathcal{T}_{\Ising} \to \mathbb{C}\}.
\end{equation}
This is also a weak Hopf algebra with the structure given by \emph{canonical pairing} $\langle \psi, h\rangle:=\psi(h)$:
\begin{align}
&	\langle \hat{\mu}(\varphi\otimes \psi),x\rangle=\langle\varphi\otimes \psi, \Delta(x)\rangle,\label{eq:MultiplicationBarA}\\
&	\langle \hat{\eta} (1),x\rangle= \varepsilon(x),   \; \text{i.e.,}\; \hat{1}=\varepsilon,\\
&   \langle \hat{\Delta}(\varphi), x\otimes y \rangle=\langle \varphi, \mu(x\otimes y)\rangle,\label{eq:ComultiplicationBarA} \\
&    \hat{\varepsilon}(\varphi)=\langle \varphi, \eta(1)\rangle,\\
&    \langle \hat{S}(\varphi),x\rangle =\langle \varphi, S(x)\rangle,
\end{align}
where $\hat{\mu}$, $\hat{\eta}$, $\hat{\Delta}$, $\hat{\varepsilon}$, $\hat{S}$ are structure maps for $\hat{\mathcal{T}}_{\Ising}$. In this work, whenever necessary, we will use hat notion to emphasize the dual structure morphisms of dual algebra, and will use Roman letter to denote element is an algebra but Greek letter to denote the elements in dual algebra.

Crucially, the characters of representations are in the $\hat{\mathcal{T}}_{\Ising}$. For objects in $\Ising$, we denote the corresponding character as $\chi_{\one},\chi_{\psi},\chi_{\sigma}$. 
The multiplication of these characters is consistent with the fusion rule of $\Ising$ in the sense that 
\begin{equation}
    \chi_a\chi_b=\sum_{c\in \Irr(\Ising)} N_{ab}^c \chi_c.
\end{equation}
This means $\sum_{(h)} \chi_{a}(h^{\cone})\otimes \chi_b(h^{\ctwo}) = \sum_{c\in \Irr(\Ising)} N_{ab}^c \chi_c(h)$.

We emphasize that the Ising boundary tube algebra $\mathcal{T}_{\mathsf{Ising}}$ is Morita equivalent, as weak Hopf algebras, to the Ising weak Hopf algebra $\mathcal{A}_{\mathsf{Ising}}$ reconstructed via Tannaka–Krein duality. A summary of their relationship is provided in Table~\ref{tab:IsingQD}.
If we use $\cA_{\Ising}$, we have $\Rep(\cA_{\Ising})\simeq \Ising$, thus, $\chi_{\one},\chi_{\psi},\chi_{\sigma}$ form a basis of $\hat{\cA}_{\Ising}$. However, as we have mentioned before, the explicit basis of ${\cA}_{\Ising}$ is difficult to write down, thus we choose the use Ising boundary tube algebra $\mathcal{T}_{\Ising}$.

An interesting property of $\mathcal{T}_{\mathsf{Ising}}$ is that its dual can also be understood from the perspective of boundary tube algebras~\cite{jia2024weakTube,jia2025weakhopftubealgebra}. This is because the category $\mathsf{Ising}$ can be regarded as a left $\mathsf{Ising}$-module category, a right $\mathsf{Ising}$-module category, or more generally as a $\mathsf{Ising}|\mathsf{Ising}$ bimodule category. While $\mathcal{T}_{\mathsf{Ising}}$ is constructed from the left module structure, we may likewise consider the corresponding construction from the right module structure and obtain $\mathbf{Tube}(\Ising_{\Ising})$ with bases:
\begin{equation}
\left\{  
\begin{aligned}
    \begin{tikzpicture}
        \begin{scope}
            \fill[gray!20]
                (0,1.1) arc[start angle=90, end angle=-90, radius=1.1] -- 
                (0,-0.5) arc[start angle=-90, end angle=90, radius=0.5] -- cycle;
        \end{scope}
           \draw[line width=0.6pt,black] (0,0.5)--(0,1.1);
        \draw[line width=.6pt,black] (0,-0.5)--(0,-1.1);
        \draw[blue, line width=0.6pt] (0,-0.8) arc[start angle=-90, end angle=90, radius=0.8];
        \node[line width=0.6pt, dashed, draw opacity=0.5] at (0,1.3) {$g$};
        \node[line width=0.6pt, dashed, draw opacity=0.5] at (0,-1.3) {$c$};
        \node[line width=0.6pt, dashed, draw opacity=0.5] at (1,0) {$b$};
        \node[line width=0.6pt, dashed, draw opacity=0.5] at (-0.3,-0.7) {$\nu$};
        \node[line width=0.6pt, dashed, draw opacity=0.5] at (0,-0.2) {$e$};
        \node[line width=0.6pt, dashed, draw opacity=0.5] at (0,0.2) {$f$};
        \node[line width=0.6pt, dashed, draw opacity=0.5] at (-0.3,0.7) {$\mu$};
    \end{tikzpicture}
\end{aligned}
:\,  \begin{aligned}
    &b,c,e,f,g\in \Irr(\Ising),\\
    &\mu \in \Hom_{\Ising}(f\otimes b,g),\\
    &\nu\in \Hom_{\Ising}(c,e\otimes b) 
\end{aligned}
\right\}.
\label{eq:tubebasis}
\end{equation}
We hereinafter denote the right-module version as $\mathcal{T}'_{\mathsf{Ising}}$. Its weak Hopf algebra structure takes a form similar to that given in Eqs.~\eqref{eq:WHA1}–\eqref{eq:WHA6}. As pointed out in~\cite{jia2025weakhopftubealgebra} (see Lemma 14 and Proposition 17 therein), $\mathcal{T}'_{\mathsf{Ising}}$ can be embedded into the dual algebra $\hat{\mathcal{T}}_{\mathsf{Ising}}$.
Since $\mathcal{T}'_{\mathsf{Ising}}$ also has dimension $34$, and its basis can be written explicitly in the same form as that of $\mathcal{T}_{\mathsf{Ising}}$, we conclude that $\mathcal{T}'_{\mathsf{Ising}} \cong \hat{\mathcal{T}}_{\mathsf{Ising}}^{\mathrm{cop}}$ (here, for a weak Hopf algebra $\cA$, $\mathcal{A}^{\mathrm{cop}}$ denotes the coopposite weak Hopf algebra).
The skew-pairing between $\mathcal{T}'_{\mathsf{Ising}}$ and $\mathcal{T}_{\mathsf{Ising}}$ is given by
\begin{equation} \label{eq:pairing_def}
\begin{aligned}
  &  p\left(\begin{aligned}
    \begin{tikzpicture}[scale=0.65]
        \begin{scope}
            \fill[gray!15]
                (0,-1.5) arc[start angle=-90, end angle=90, radius=1.5] -- 
                (0,0.5) arc[start angle=90, end angle=-90, radius=0.5] -- cycle;
        \end{scope}
        \draw[line width=0.6pt,black] (0,0.5)--(0,1.5);
        \draw[line width=.6pt,black] (0,-0.5)--(0,-1.5);
        \draw[blue, line width=0.6pt] (0,1.1) arc[start angle=90, end angle=-90, radius=1.1];
        \node[ line width=0.6pt, dashed, draw opacity=0.5] (a) at (0,-1.7){$\scriptstyle s$};
        \node[ line width=0.6pt, dashed, draw opacity=0.5] (a) at (1.3,0){$\scriptstyle b$};
        \node[ line width=0.6pt, dashed, draw opacity=0.5] (a) at (0.25,-0.7){$\scriptstyle t$};
        \node[ line width=0.6pt, dashed, draw opacity=0.5] (a) at (-0.25,-1.15){$\scriptstyle \mu$};
        \node[ line width=0.6pt, dashed, draw opacity=0.5] (a) at (0.25,0.7){$\scriptstyle u$};
        \node[ line width=0.6pt, dashed, draw opacity=0.5] (a) at (0,1.7){$\scriptstyle v$};
        \node[ line width=0.6pt, dashed, draw opacity=0.5] (a) at (-0.25,1.15){$\scriptstyle \gamma$};
    \end{tikzpicture}
\end{aligned}\;,\;
\begin{aligned}
    \begin{tikzpicture}[scale=0.65]
        \begin{scope}
            \fill[gray!15]
                (0,1.5) arc[start angle=90, end angle=270, radius=1.5] -- 
                (0,-0.5) arc[start angle=270, end angle=90, radius=0.5] -- cycle;
        \end{scope}
        \draw[line width=0.6pt,black] (0,0.5)--(0,1.5);
        \draw[line width=.6pt,black] (0,-0.5)--(0,-1.5);
        \draw[red, line width=0.6pt] (0,1.1) arc[start angle=90, end angle=270, radius=1.1];
        \node[ line width=0.6pt, dashed, draw opacity=0.5] (a) at (-1.3,0){$\scriptstyle a$};
        \node[ line width=0.6pt, dashed, draw opacity=0.5] (a) at (0,-1.7){$\scriptstyle x$};
        \node[ line width=0.6pt, dashed, draw opacity=0.5] (a) at (-0.25,-0.7){$\scriptstyle y$};
        \node[ line width=0.6pt, dashed, draw opacity=0.5] (a) at (0.26,-1.15){$\scriptstyle \nu$};
        \node[ line width=0.6pt, dashed, draw opacity=0.5] (a) at (-0.25,0.7){$\scriptstyle z$};
        \node[ line width=0.6pt, dashed, draw opacity=0.5] (a) at (0,1.7){$\scriptstyle w$};
        \node[ line width=0.6pt, dashed, draw opacity=0.5] (a) at (0.2,1.15){$\scriptstyle \zeta$};
    \end{tikzpicture}
\end{aligned}
    \right)  \\
     = & \frac{\delta_{s,w}\delta_{t,x}\delta_{u,y}\delta_{v,z}}{d_s}\;\begin{aligned}
        \begin{tikzpicture}[scale=0.7]
        \path[black!60, fill=gray!15] (0,0) circle[radius=1.4];
             \draw[line width=.6pt,black] (0,-1.4)--(0,1.4);
             \draw[red, line width=0.6pt] (0,0.95) arc[start angle=90, end angle=270, radius=0.7];
             \draw[blue, line width=0.6pt] (0,-0.95) arc[start angle=-90, end angle=90, radius=0.7];
             \draw[line width=.6pt,black] (0,1.4) arc[start angle=90, end angle=-90, radius=1.4];
             \node[ line width=0.6pt, dashed, draw opacity=0.5] (a) at (1.2,0){$\scriptstyle \bar{s}$};
            \node[ line width=0.6pt, dashed, draw opacity=0.5] (a) at (-0.9,0.4){$\scriptstyle a$};
            \node[ line width=0.6pt, dashed, draw opacity=0.5] (a) at (0.9,-0.4){$\scriptstyle b$};
            \node[ line width=0.6pt, dashed, draw opacity=0.5] (a) at (-0.2,-0.7){$\scriptstyle t$};
            \node[ line width=0.6pt, dashed, draw opacity=0.5] (a) at (-0.2,-1.1){$\scriptstyle \mu$};
            \node[ line width=0.6pt, dashed, draw opacity=0.5] (a) at (0.2,-0.5){$\scriptstyle \nu$};
            \node[ line width=0.6pt, dashed, draw opacity=0.5] (a) at (-0.2,-0.1){$\scriptstyle u$};
            \node[ line width=0.6pt, dashed, draw opacity=0.5] (a) at (-0.2,0.7){$\scriptstyle v$};
            \node[ line width=0.6pt, dashed, draw opacity=0.5] (a) at (-0.2,0.3){$\scriptstyle \gamma$};
            \node[ line width=0.6pt, dashed, draw opacity=0.5] (a) at (0.2,1){$\scriptstyle \zeta$};
        \end{tikzpicture}
    \end{aligned}\;. 
\end{aligned}
\end{equation}
It was proved in \cite{jia2025weakhopftubealgebra} that this defines a skew-pairing. Furthermore, the right-hand side of Eq.~\eqref{eq:pairing_def} can be explicitly evaluated using topological local move to yield a complex number.

\subsection{Quantum double of Ising boundary tube algebra}

Within the SymTFT framework for the Ising fusion category symmetry $\Ising$, the associated one-higher-dimensional bulk is characterized by the topological order $\mathcal{Z}(\Ising)$, namely the Drinfeld center (or quantum double) of $\Ising$. Reformulated in the language of weak Hopf algebra symmetry, one considers the weak Hopf algebra $\mathcal{T}_{\mathsf{Ising}}$ corresponding to the Ising category, and defines its quantum double $D(\mathcal{T}_{\mathsf{Ising}})$ using both $\mathcal{T}_{\mathsf{Ising}}$ and its dual $\widehat{\mathcal{T}}_{\mathsf{Ising}}$.
It is known that there exists a categorical equivalence
\[
\mathcal{Z}(\Ising) \simeq \mathcal{Z}(\Rep(\mathcal{T}_{\mathsf{Ising}})) \simeq \Rep(D(\mathcal{T}_{\mathsf{Ising}})),
\]
indicating that the bulk topological order is fully captured by the representation category of the quantum double $D(\mathcal{T}_{\mathsf{Ising}})$. This structure serves as the weak Hopf charge symmetry in the bulk.
Therefore, the bulk is effectively characterized by the quantum double $D(\mathcal{T}_{\mathsf{Ising}})$.

Since the quantum double plays a central role in this work, we briefly review its definition here.
Let $\mathcal{A}$ be a weak Hopf algebra. Its dual $(\hat{\mathcal{A}}, \hat{\mu}, \hat{\eta}, \hat{\Delta}, \hat{\epsilon}, \hat{S})$ is also a weak Hopf algebra. We may consider its coopposite algebra $\hat{\mathcal{A}}^{\mathrm{cop}} := (\hat{\mathcal{A}}, \hat{\mu}, \hat{\eta}, \hat{\Delta}^{\mathrm{op}}, \hat{\epsilon}, \hat{S}^{-1})$, where $\hat{\Delta}^{\mathrm{op}} := \tau \circ \hat{\Delta}$, with $\tau(\psi \otimes \phi) = \phi \otimes \psi$ the swap map.

We denote the canonical pairing between $\hat{\mathcal{A}}$ and $\mathcal{A}$ by $\langle \psi, h \rangle := \psi(h)$. When regarded as a pairing between $\hat{\mathcal{A}}^{\mathrm{cop}}$ and $\mathcal{A}$, it defines a skew-pairing.
The tensor product algebra $\hat{\mathcal{A}}^{\mathrm{cop}} \otimes \mathcal{A}$ is equipped with a multiplication defined as follows~\cite{drinfel1988quantum,majid1990physics,majid1994some,nikshych2002finite}:
\begin{equation}
\begin{aligned}
       & (\varphi \otimes h)(\psi \otimes g) \\
    = &\sum_{(h)} \sum_{(\psi)} \varphi \psi^{(2)} \otimes h^{(2)} g \, \langle \psi^{(1)}, S^{-1}(h^{(3)}) \rangle \langle \psi^{(3)}, h^{(1)} \rangle.
\end{aligned}
\end{equation}
It is useful to express this multiplication via the so-called straightening relation~\cite{kassel2012quantum}:
\begin{equation}
    h \cdot \psi 
    = \psi^{(2)} \otimes h^{(2)} \, \langle \psi^{(1)}, S^{-1}(h^{(3)}) \rangle \langle \psi^{(3)}, h^{(1)} \rangle.
\end{equation}
The unit element of this algebra is $\varepsilon \otimes 1_{\mathcal{A}}$.

Unlike in the case of Hopf algebras (which can be viewed as a special subclass of weak Hopf algebras), for a general weak Hopf algebra, we must consider a nontrivial two-sided ideal \(M\) in \(\hat{\mathcal{A}}^{\mathrm{cop}} \otimes \mathcal{A}\), which is spanned by elements of the form
\begin{align}
    \varphi \otimes xh - \varphi (x \rightharpoonup \varepsilon) \otimes h, \quad &x \in \mathcal{A}_L, \\
    \varphi \otimes yh - \varphi (\varepsilon \leftharpoonup y) \otimes h, \quad &y \in \mathcal{A}_R,
\end{align}
where \(\mathcal{A}_L\) and \(\mathcal{A}_R\) denote the images of the source and target counital maps, respectively.
We define the quantum double of \(\mathcal{A}\), denoted by \(D(\mathcal{A})\), as the quotient algebra
\[
D(\mathcal{A}) := (\hat{\mathcal{A}}^{\mathrm{cop}} \otimes \mathcal{A}) / M,
\]
with equivalence classes represented as \([\varphi \otimes h]\), for \(\varphi \otimes h \in \hat{\mathcal{A}}^{\mathrm{cop}} \otimes \mathcal{A}\).

The quantum double of a weak Hopf algebra \(\cA\), denoted by \(D(\cA)\), is a weak Hopf algebra equipped with the following structure:

\begin{enumerate}
    \item[(1)] The multiplication operation is defined by
    \[
    \begin{aligned}
         &[\varphi \otimes h] [\psi \otimes g] \\
         = &\sum_{(\psi), (h)} [\varphi \psi^{(2)} \otimes h^{(2)} g] \langle \psi^{(1)}, S^{-1}(h^{(3)}) \rangle \langle \psi^{(3)}, h^{(1)} \rangle.
    \end{aligned}
    \]
    
    \item[(2)] The unit element is
    \[
    [\varepsilon \otimes 1_{\cA}].
    \]
    
    \item[(3)] The comultiplication is given by
    \[
    \Delta([\varphi \otimes h]) = \sum_{(\varphi), (h)} [\varphi^{(2)} \otimes h^{(1)}] \otimes [\varphi^{(1)} \otimes h^{(2)}].
    \]
    
    \item[(4)] The counit is defined as
    \[
    \varepsilon ([\varphi \otimes h]) = \langle \varphi, \varepsilon_R(S^{-1}(h)) \rangle.
    \]
    
    \item[(5)] The antipode is given by
    \[
    \begin{aligned}
          & S([\varphi \otimes h]) \\
          = & \sum_{(\varphi), (h)} [\hat{S}^{-1}(\varphi^{(2)}) \otimes S(h^{(2)})] \langle \varphi^{(1)}, h^{(3)} \rangle \langle \varphi^{(3)}, S^{-1}(h^{(1)}) \rangle.
    \end{aligned}
    \]
\end{enumerate}

To make the model more calculable, here we still consider the Ising boundary tube algebra $\mathcal{T}_{\Ising}$. Its quantum double has a nice property that makes it more calculable, since the quantum double $D(\mathcal{T}_{\Ising})$ can be regarded as a domain wall tube algebra $\Tube(_{\Ising}\Ising_{{\Ising}})$ constructed for bimodule category $_{\Ising}\Ising_{{\Ising}}$ \cite{jia2025weakhopftubealgebra}:
\begin{equation}
    D(\mathcal{T}_{\Ising})\cong \Tube(_{\Ising}\Ising_{{\Ising}})
\end{equation}
Thus, we can focus on the domain wall tube algebra $\Tube(_{\Ising}\Ising_{{\Ising}})$, its basis is given by:
\begin{equation} \label{eq:tube_basis}
 \left\{  
   \begin{aligned}

    \end{aligned}
   \;:\; 
   \begin{aligned}
     & a,b, x,y,z,u,v,w\in \Irr(\Ising), \\
     & \mu\in\Hom_{\Ising}(x,y\otimes b),\\
     &\nu\in\Hom_{\Ising}(y,a\otimes z),\\
     & \zeta\in\Hom_{\Ising}(a\otimes u,v),\\
     &\gamma\in \Hom_{\Ising}(v\otimes b,w)
   \end{aligned}
    \right\}.
\end{equation}
The weak Hopf structure is discussed in detail in Refs~\cite{jia2024weakTube,jia2025weakhopftubealgebra}.

The unit element is defined by 
\begin{equation} \label{eq:bi_tube_unit}
    1=\sum_{x,z\in\Irr(\Ising)} \;\begin{aligned}
        %
    \end{aligned}\;\;. 
\end{equation}

The multiplication map $\mu$ is given by 
\begin{align}
     & \mu\left( \begin{aligned}
        %
\caption{\label{tab:IsingQD} The Ising weak Hopf algebra obtained via Tannaka–Krein reconstruction and the Ising boundary tube algebra are Morita equivalent as weak Hopf algebras; both can be used to construct the Ising cluster state model.}
\end{table*}

 The counit is defined as 
\begin{equation} \label{eq:bi_tube_counit}
\begin{aligned}
       &  \varepsilon\left(\begin{aligned}%
\end{aligned}\\
=&\delta_{z,u}\delta_{x,w}\delta_{y,v}\delta_{\nu,\zeta}\delta_{\mu,\gamma} \sqrt{\frac{d_a d_zd_b}{d_x}}
\end{aligned}
\end{equation}

The comultiplication (or coproduct) is defined as 
\begin{equation}\label{eq:bi_tube_coprod}
    \begin{aligned} 
   &  \Delta\left(\begin{aligned}%
\end{aligned}\;\;.
\end{aligned}
\end{equation}

The antipode is defined as 
\begin{equation} \label{eq:bi_tube_antipode}
    S\left( 
          \begin{aligned}%
    \end{aligned}\;\;. 
\end{equation}

Hereinafter, for notational convenience, we will not distinguish between $D(\Tsing)$ and $\Tube(_{\Ising}\Ising_{\Ising})$, nor between $\mathcal{T}'_{\Ising}$ and $\hat{\mathcal{T}}_{\Ising}$.

Both of $\Tsing$ and $\hat{\mathcal{T}}_{\Ising}$ can be embedded into $D(\Tsing)$  with the respective embedding given by
\begin{equation}
    \begin{aligned}
    %
    \end{aligned},
\end{equation}
it is straightforward to verify that the above maps are weak Hopf algebra maps.

\section{Cluster state model with Ising fusion category symmetry} 
\label{sec:IsingCluHam}
The weak Hopf SymTFT is outlined in \cite{jia2024generalized,jia2024weakhopfnoninvertible}, here we apply this framework to the Ising weak Hopf symmetry $\mathcal{T}_{\Ising}$. The bulk phase is now characterized by weak Hopf quantum double model with weak Hopf symmetry given by $\mathcal{T}_{\Ising}$.

\subsection{Ising weak Hopf SymTFT}

Recall that the SymTFT construction for a $(1+1)$D theory is characterized by a $(2+1)$D topological order defined on a sandwich manifold $\mathbb{M}^1 \times [0,1]$, see Figure~\ref{fig:SymTFT}. The boundary at $\mathbb{M}^1 \times \{0\}$ hosts a symmetry boundary $\mathcal{B}_{\rm sym}$, while the boundary at $\mathbb{M}^1 \times \{1\}$ carries a physical boundary $\mathcal{B}_{\rm phys}$.
In our case, the symmetry boundary is given by $\mathcal{B}_{\rm sym} = \Ising$, and the bulk topological order is determined by the Drinfeld center of $\Ising$. Since $\Ising$ is a modular tensor category, its Drinfeld center takes the form
\begin{equation}
    \mathcal{Z}(\Ising) = \Ising \boxtimes \overline{\Ising},
\end{equation}
where $\overline{\Ising}$ denotes the fusion category with the same fusion rules as $\Ising$ but equipped with the reversed braiding, i.e., $\bar{\beta}_{a,b} = \beta_{a,b}^{-1}$, and with complex-conjugated topological spins, $\bar{\theta}_a = \theta_a^{-1}$ (this is a result of the reversed braiding).
The bulk anyon type is thus described by $\{a \boxtimes b \mid a, b \in \Ising\}$.
The quantum dimension is given by $d_{a\boxtimes b}=d_ad_b$ and topological spin is given by $\theta_{a\boxtimes b}=\theta_a\bar{\theta}_b$, see Table~\ref{tab:IsingDoubleAnyon}. 
Consider the central charge
\begin{equation}
\begin{aligned}
       e^{2 \pi i c/8} = \frac{1}{\operatorname{FPdim} \Ising } \sum_{a \in \mathrm{Irr}(\Ising)} d_a^2 \theta_a, \\
 e^{2 \pi i \bar{c}/8} =\frac{1}{\operatorname{FPdim} \Ising }  \sum_a d_a^2 \bar{\theta}_a= e^{-2 \pi i c/8}. 
\end{aligned}
\end{equation}
The full theory is non-chiral, since the chiral central charge cancels: $c_{\Ising \boxtimes \overline{\Ising}} = \frac{1}{2} - \frac{1}{2} = 0 \,\, \operatorname{mod} 8$.

\begin{table*}
\caption{\label{tab:IsingDoubleAnyon}Anyons in the $\Ising \boxtimes \overline{\Ising}$ theory, their quantum dimensions, and topological spins.}
\centering
\renewcommand{\arraystretch}{1.3}
\begin{tabular}{l|c|c|c|c|c|c|c|c|c}
\hline
\hline
Anyon & $\mathbb{1} \boxtimes \mathbb{1}$ & $\psi \boxtimes \mathbb{1}$ & $\sigma \boxtimes \mathbb{1}$ & $\mathbb{1} \boxtimes \psi$ & $\psi \boxtimes \psi$ & $\sigma \boxtimes \psi$ & $\mathbb{1} \boxtimes \sigma$ & $\psi \boxtimes \sigma$ & $\sigma \boxtimes \sigma$ \\
\hline
FPdim & $1$ & $1$ & $\sqrt{2}$ & $1$ & $1$ & $\sqrt{2}$ & $\sqrt{2}$ & $\sqrt{2}$ & $2$ \\
\hline
Topopoligical spin & $1$ & $-1$ & $e^{\pi i/8}$ & $-1$ & $1$ & $-e^{\pi i/8}$ & $e^{-\pi i/8}$ & $-e^{-\pi i/8}$ & $1$ \\
\hline
\hline
\end{tabular}
\end{table*}

Our central idea is to construct a weak Hopf quantum double model based on $\mathcal{T}_{\Ising}$, and then place it on a sandwich manifold. 
In the bulk, the corresponding quantum double model exhibits a topological order characterized by the representation category of the quantum double of $\mathcal{T}_{\Ising}$. As discussed earlier, the quantum double of the Ising tube algebra is equivalent to the domain wall Ising tube algebra $\mathbf{Tube}(_{\Ising}\Ising_{\Ising})$. 
In summary, we have the equivalence:
\begin{equation}
\begin{aligned}
     & \mathcal{Z}(\Ising) = \Ising \boxtimes \overline{\Ising} \simeq \Rep(D(\mathcal{T}_{\Ising})) \\
      &\simeq \Rep(\mathbf{Tube}(_{\Ising}\Ising_{\Ising})).
\end{aligned}
\end{equation}
Hence, the weak Hopf quantum double model constructed from $\mathcal{T}_{\Ising}$ realizes the desired bulk topological order.

The SymTFT framework provides a powerful tool to understand non-invertible symmetries. For the $\Ising$ SymTFT, we take the symmetry boundary to be $\Ising$, while the bulk realizes the doubled Ising phase. 
To determine the topological boundary condition of Ising quantum double model, we need to determine the Lagrangian algebras in bulk phase $\Ising \boxtimes \overline{\Ising}$, notice that 
\begin{equation}
    \operatorname{FPdim} \Ising \boxtimes \overline{\Ising} = \sum_{a,b\in \Ising} (d_ad_b)^2=16.
\end{equation}
The Lagrangian algebra $\mathcal{L}$ must have quantum dimension $d_{\mathcal{L}}=4$.
Since $\mathcal{L}=(\one\boxtimes \one ) \oplus X$ with $X$ does not contain vacuum charge.
The symmetry boundary $\mathcal{B}_{\mathrm{sym}} = \Ising$ corresponds to a smooth boundary characterized by the Lagrangian algebra given by the diagonal object
\begin{equation}
    \mathcal{L}_s = (\mathbb{1} \boxtimes \mathbb{1}) \oplus (\psi \boxtimes \psi) \oplus (\sigma \boxtimes \sigma).
\end{equation}
Condensing this Lagrangian algebra yields a boundary topological phase described by the category of $\mathcal{L}_s$-modules in $\mathcal{Z}(\Ising)$, which is equivalent to $\Ising$, viz. 
$\mathcal{Z}(\Ising)_{\mathcal{L}_s} \simeq \Ising$.

\subsection{Ising weak Hopf cluster state model}

We now construct the Ising weak Hopf cluster state model, which can be viewed as an ultra-thin Ising quantum double model with one smooth boundary (the symmetry boundary) and one rough boundary (the physical boundary).\footnote{Here, the notion of a rough boundary refers to the removal of degrees of freedom at that boundary, which is commonly used in lattice model since the original work of Ref.~\cite{bravyi1998quantum}. This differs from another common usage in the literature, where the rough boundary corresponds to the module category $\Vect$ over the input fusion category in the string-net model framework, or equivalently, the comodule algebra $\mathbb{C}$ over the input weak Hopf algebra in the quantum double model framework. Since $\Ising$ is an anomalous symmetry, $\Vect$ is not a module category over $\Ising$, and $\mathbb{C}$ is not a comodule algebra over $\mathcal{T}_{\Ising}$. But for anomaly-free symmetry, two notions are the same.}
We argue that this is a natural generalization of cluster state model for any non-invertible symmetries \cite{jia2024generalized,jia2024weakhopfnoninvertible}.

The model is defined on an ultra-thin sandwich lattice, which can be regarded as a weak Hopf quantum double model with two boundaries~\cite{Jia2023weak}:
\begin{equation*}
\begin{aligned}
\begin{tikzpicture}
    \def\n{5}
    \def\s{1}   
    \fill[yellow!10] (0, 0) rectangle (\n*\s+\s, \s);

    \foreach \i in {0,...,\n} {
        \draw[-stealth, line width=1pt, red, midway] (\i*\s, 0) -- (\i*\s+\s, 0);
        \draw[-stealth, line width=1pt, blue, midway] (\i*\s, \s) -- (\i*\s+\s, \s);
        \draw[-stealth, line width=1pt, midway] (\i*\s, 0) -- (\i*\s, \s);
    }
    \draw[-stealth, line width=1pt, midway] (\n*\s+\s, 0) -- (\n*\s+\s, \s);
    \draw[-stealth, white, line width=2pt, midway] (6, 0) -- (6, 1.02);
\end{tikzpicture}
\end{aligned}
\end{equation*}
Periodic or open boundary conditions can be imposed along the horizontal direction. 
The bulk edges are drawn in black, with the associated edge Hilbert space taken as the Ising weak Hopf qubit $\mathcal{H}_e = \mathcal{T}_{\Ising}$. 
The red edges along the symmetry (smooth) boundary correspond to a left $\mathcal{T}_{\Ising}$-comodule algebra $\mathcal{K}_s = \mathcal{T}_{\Ising}$. 
The blue edges along the physical boundary are associated with an arbitrary right $\mathcal{T}_{\Ising}$-comodule algebra $\mathcal{K}_p$, assuming the model is gapped.
This setup ensures that the total Hilbert space admits a well-defined tensor product structure. The construction of the lattice model follows the framework developed in Ref.~\cite{jia2024weakhopfnoninvertible}.

In this work, we focus on the cluster state model, where the physical boundary edges are entirely removed:
\begin{equation*}
\begin{aligned}
\begin{tikzpicture}
    \def\n{5}
    \def\s{1}   
    \fill[yellow!10] (0, 0) rectangle (\n*\s+\s, \s);

    \foreach \i in {0,...,\n} {
        \draw[-stealth, line width=1pt, red, midway] (\i*\s, 0) -- (\i*\s+\s, 0);
        \draw[-stealth, line width=1pt, midway] (\i*\s, 0) -- (\i*\s, \s);
    }
    \draw[-stealth, line width=1pt, midway] (\n*\s+\s, 0) -- (\n*\s+\s, \s);
    \draw[-stealth, white, line width=2pt, midway] (6, 0) -- (6, 1.02);
\end{tikzpicture}
\end{aligned}
\end{equation*}
In this configuration, the Hamiltonian consists solely of bulk face terms and vertex terms located on the symmetry boundary.

To construct the lattice model with Ising fusion category symmetry, we begin by introducing the generalized Pauli operators. The regular action of the Ising weak Hopf algebra $\mathcal{T}_{\Ising}$ on itself can be viewed as a generalization of Pauli $X$-type operators. 
For the left action \( \mathcal{T}_{\Ising} \curvearrowright \mathcal{T}_{\Ising} \), we define:
\begin{equation}
\XR_g |h\rangle = |gh\rangle, \quad \XL_g |h\rangle = |hS^{-1}(g)\rangle, \quad \forall\, g, h \in \mathcal{T}_{\Ising}.
\end{equation}
The dual space \( \hat{\mathcal{T}}_{\Ising} := \Hom(\mathcal{T}_{\Ising}, \mathbb{C}) \) also forms a weak Hopf algebra. As previously noted, $\hat{\mathcal{T}}_{\Ising}$ coincides with the right boundary tube algebra $\mathcal{T}'_{\Ising} = \mathbf{Tube}(\Ising_{\Ising})$, more precisely,
$\hat{\mathcal{T}}_{\Ising} = (\mathcal{T}'_{\Ising})^{\mathrm{cop}}$.

There are also canonical actions of the dual weak Hopf algebra $\hat{\mathcal{T}}_{\Ising}$ on the Hopf qudit $\mathcal{T}_{\Ising}$. 
Based on the pairing defined in Eq.~\eqref{eq:pairing_def}, and using Sweedler’s notation, we define the actions as follows:
\begin{equation}
\varphi \rightharpoonup x := \sum_{(x)} x^{(1)} \langle \varphi, x^{(2)} \rangle, \quad
x \leftharpoonup \varphi := \sum_{(x)} \langle \varphi, x^{(1)} \rangle x^{(2)},
\end{equation}
for all \( \varphi \in \hat{\mathcal{T}}_{\Ising} \) and \( x \in \mathcal{T}_{\Ising} \).
For the left action \( \hat{\mathcal{T}}_{\Ising} \curvearrowright \mathcal{T}_{\Ising} \), we define:
\begin{equation}
\ZR_{\psi} |h\rangle = |\psi \rightharpoonup h\rangle = \sum_{(h)} \psi(h^{\ctwo})\, |h^{\cone}\rangle,
\end{equation}
\begin{equation}
\ZL_{\psi} |h\rangle = |h \leftharpoonup \hat{S}(\psi)\rangle = \sum_{(h)} \psi(S(h^{\cone}))\, |h^{\ctwo}\rangle.
\end{equation}
These operators can be viewed as generalized Pauli \( Z \)-type operators.

Our model depends on a special element in weak Hopf algebra, called Haar integral, by definition, it's an element $\lambda$ that $\lambda g=\lambda \varepsilon_R(g)$, $g\lambda=\varepsilon_L(g)\lambda$ for all $g$ in the algebra and $\varepsilon_L(\lambda)=\varepsilon_R(\lambda)=1$ thus $\lambda$ is an idempotent $\lambda^2=\lambda$.

It can be proved that the boundary tube algebra is a $C^*$ weak Hopf algebra \cite{Kitaev2012boundary, bridgeman2023invertible, jia2024weakTube, jia2025weakhopftubealgebra}, and therefore it admits a Haar integral of the form
\begin{equation}\label{eq:HaarTube}
    \lambda = \frac{1}{\operatorname{rank} \eC} \sum_{a,x,y,\mu} 
    \sqrt{\frac{d_a}{d_x^3 d_y}}
        \begin{aligned}
    \begin{tikzpicture}
        \begin{scope}
            \fill[gray!20]
                (0,1.1) arc[start angle=90, end angle=270, radius=1.1] -- 
                (0,-0.5) arc[start angle=270, end angle=90, radius=0.5] -- cycle;
        \end{scope}
           \draw[line width=0.6pt,black] (0,0.5)--(0,1.1);
        \draw[line width=.6pt,black] (0,-0.5)--(0,-1.1);
        \draw[red, line width=0.6pt] (0,0.8) arc[start angle=90, end angle=270, radius=0.8];
        \node[line width=0.6pt, dashed, draw opacity=0.5] at (0,1.3) {$y$};
        \node[line width=0.6pt, dashed, draw opacity=0.5] at (0,-1.3) {$y$};
        \node[line width=0.6pt, dashed, draw opacity=0.5] at (-1,0) {$a$};
        \node[line width=0.6pt, dashed, draw opacity=0.5] at (0.3,-0.7) {$\mu$};
        \node[line width=0.6pt, dashed, draw opacity=0.5] at (0,-0.3) {$x$};
        \node[line width=0.6pt, dashed, draw opacity=0.5] at (0,0.3) {$x$};
        \node[line width=0.6pt, dashed, draw opacity=0.5] at (0.3,0.7) {$\mu$};
    \end{tikzpicture}
\end{aligned}.
\end{equation}
A detailed proof can be found in Ref.~\cite{jia2025weakhopftubealgebra}. In the case where $\eC = \Ising$, we have $\operatorname{FPdim} \Ising = 4$ and $\operatorname{rank} \Ising = 3$. 

Via the identification of the dual of $\TIsing$ with $(\mathcal{T}'_{\Ising})^{\mathrm{cop}} = \mathbf{Tube}(\Ising_{\Ising})^{\mathrm{cop}}$, the Haar measure (i.e., the Haar integral of the dual algebra) is given by
\begin{equation}\label{eq:HaarMTube}
    \Lambda = \frac{1}{\operatorname{rank} \eC} \sum_{a,x,y,\mu} 
    \sqrt{\frac{d_a}{d_x^3 d_y}}
        \begin{aligned}
    \begin{tikzpicture}
        \begin{scope}
            \fill[gray!20]
                (0,1.1) arc[start angle=90, end angle=-90, radius=1.1] -- 
                (0,-0.5) arc[start angle=-90, end angle=90, radius=0.5] -- cycle;
        \end{scope}
           \draw[line width=0.6pt,black] (0,0.5)--(0,1.1);
        \draw[line width=.6pt,black] (0,-0.5)--(0,-1.1);
        \draw[blue, line width=0.6pt] (0,-0.8) arc[start angle=-90, end angle=90, radius=0.8];
        \node[line width=0.6pt, dashed, draw opacity=0.5] at (0,1.3) {$y$};
        \node[line width=0.6pt, dashed, draw opacity=0.5] at (0,-1.3) {$y$};
        \node[line width=0.6pt, dashed, draw opacity=0.5] at (1,0) {$a$};
        \node[line width=0.6pt, dashed, draw opacity=0.5] at (-0.3,-0.7) {$\mu$};
        \node[line width=0.6pt, dashed, draw opacity=0.5] at (0,-0.3) {$x$};
        \node[line width=0.6pt, dashed, draw opacity=0.5] at (0,0.3) {$x$};
        \node[line width=0.6pt, dashed, draw opacity=0.5] at (-0.3,0.7) {$\mu$};
    \end{tikzpicture}
\end{aligned}.
\end{equation}
The Haar measure induces an inner product on $\TIsing$ defined by
\begin{equation}
    \langle x, y \rangle = \Lambda(x^* y), \quad x, y \in \TIsing,
\end{equation}
where the $*$-operation $x^*$ is given in Eq.~\eqref{eq:WHA6}, the product $x^* y$ refers to the multiplication in $\TIsing$, and the evaluation is defined via the pairing in Eq.~\eqref{eq:pairing_def}.
We apply this inner product to each edge space of the cluster state lattice, where all edges are assigned the Hilbert space $\mathcal{H}_e = \TIsing$.

On the symmetry boundary, the boundary vertex operator is defined by
\begin{equation}
    \begin{aligned}
        \begin{tikzpicture}
            \fill[yellow!10] (0, 0) rectangle (2,1); 
            \draw[-stealth,blue,line width = 1.6pt] (0,0) -- (1,0); 
            \draw[-stealth,blue,line width = 1.6pt] (1,0) -- (2,0); 
            \draw[-stealth,black,line width=1.0pt] (1,0) -- (1,1); 
            \draw[cyan,dotted,line width=1pt] (1,0) -- (0.5,0.5); 
            \node at (1.8,0.3) {\( y \)};
            \node at (0.2,0.3) {\( x \)};
            \node at (0.7,0.8) {\( h \)};
            \node at (1.8,-0.3) {\textcircled{2}};
            \node at (0.3,-0.3) {\textcircled{1}};
            \node at (1,1.3) {\textcircled{3}};
        \end{tikzpicture}
    \end{aligned}
    \quad 
    \begin{aligned}
        \Av(s) = \XR_{h^{\cone}} \otimes \XL_{h^{\ctwo}} \otimes \XL_{h^{\cthree}}.
    \end{aligned}
    \label{eq:Avs}
\end{equation}
More precisely,
\begin{equation}
     \begin{aligned}
        & \Av(s)\,|x, y, h\rangle \\
        &= |h^{\cone} x,\, y S^{-1}(h^{\ctwo}),\, h S^{-1}(h^{\cthree})\rangle.
    \end{aligned}
\end{equation}
By setting \( h = \lambda \), where \( \lambda \) is the Haar integral of \( \TIsing \) (see Eq.~\eqref{eq:HaarTube}), we obtain the vertex stabilizers \( \Av_{v_s} \) on the symmetry boundary.
The cocommutativity of \( \lambda \) ensures that the operator \( \Av_{v_s} \) is independent of the choice of the starting edge \( s = (v_s, f) \), where \( f \) denotes a face adjacent to the vertex \( v_s \). Consequently, the operator \( \Av_{v_s} \) depends only on the vertex \( v_s \) itself.

The face operator is constructed from generalized Pauli \( Z \) operators as follows:
\begin{equation}
\begin{aligned}
    \begin{tikzpicture}
    \begin{scope}[rotate=-90] 
        \fill[yellow!10] (0, 0) rectangle (1,1); 
        \draw[cyan,dotted,line width=1pt] (0,0) -- (0.5,0.5); 
        \draw[-stealth,red,line width=1.6pt] (1,0) -- (1,1);
        \draw[-stealth,black,line width=1.0pt] (1,1) -- (0,1); 
        \draw[-stealth,black,line width=1.0pt] (1,0) -- (0,0); 
        \node at (1.4,0.5) {\( x \)};
        \node at (0.5,1.25) {\( h \)};
        \node at (0.5,-0.25) {\( g \)};
        \node at (0.8,0.5) {\textcircled{2}};
        \node at (0.5,0.8) {\textcircled{3}};
        \node at (0.5,0.2) {\textcircled{1}};
    \end{scope}
\end{tikzpicture}   
\end{aligned}
\begin{aligned}
   \Bf^{\varphi}(s) = \ZR_{\varphi^{\cone}} \otimes \ZL_{\varphi^{\ctwo}} \otimes \ZL_{\varphi^{\cthree}}.
\end{aligned}
\end{equation}
More precisely
\begin{equation}
    \begin{aligned}
      & \Bf^{\varphi}(s)\,|g, x, h\rangle \\
       & = \sum \varphi\left(g^{\ctwo} S(x^{(1)}) S(h^{(1)}) \right) |g^{\cone},\, x^{(2)},\, h^{\ctwo}\rangle.
  \end{aligned}
\end{equation}
By setting \( \varphi = \Lambda \in \hTIsing \), the Haar measure, we obtain the face stabilizer operator \( \Bf_f = \Bf_f^{\Lambda} \). Similar to the vertex operator, the cocommutativity of \( \Lambda \) ensures that \( \Bf_f \) is independent of the choice of starting site \( s \), and thus depends only on the face \( f \).

The Hamiltonian of the Ising cluster ladder model is given by
\begin{equation}\label{eq:SymHam}
    \Hbb_{\rm Ising} = -\sum_{v_s:\, \text{sym bd}} \Av_{v_s} - \sum_f \Bf_f.
\end{equation}
The ground state of the model is called Ising cluster state.

It can be shown that all local operators commute with each other~\cite{Jia2023weak,jia2024weakhopfnoninvertible,chang2014kitaev}. On a fixed site $s = (v_s, f)$, there is a representation of the quantum double
$D(\TIsing) = \hTIsing^{\mathrm{cop}} \Join \TIsing \cong \mathbf{Tube}({}_{\Ising}\Ising_{\Ising})$,
given by:
\begin{equation}
\begin{aligned}
    \begin{tikzpicture}
        \fill[yellow!10] (0, 0) rectangle (2,1); 
        \draw[-stealth,blue,line width=1.6pt] (0,0) -- (1,0); 
        \draw[-stealth,blue,line width=1.6pt] (1,0) -- (2,0); 
        \draw[-stealth,black,line width=1.0pt] (1,0) -- (1,1); 
        \draw[-stealth,black,line width=1.0pt] (0,0) -- (0,1); 
        \draw[cyan,dotted,line width=1pt] (1,0) -- (0.5,0.5); 
        \node at (0.5,0.3) {$\Bf_f^{\psi}$};
        \node at (1.0,-0.3) {$\Av_{v_s}^{h}$};
    \end{tikzpicture}
\end{aligned}
\quad
\begin{aligned}
    & (h, \psi) \mapsto D^{(\psi \otimes h)}(s) := \Bf^{\psi}(s)\, \Av^h(s).
\end{aligned}
\label{eq:Avs}
\end{equation}
See \cite{Jia2023weak} for a proof.
Note that this operator acts nontrivially only on the four edges adjacent to the site $s$.

We can also modify the physical boundary to obtain additional models with $\Ising$ symmetry, as outlined in Ref.~\cite{jia2024weakhopfnoninvertible}. By choosing different comodule algebras $\mathcal{J}$ over the Ising weak Hopf algebra, one can realize distinct $\Ising$-symmetric phases.

\subsection{Symmetry of the Ising cluster state model}

To determine the symmetry of the Ising cluster state model, we deform the ladder (with periodic boundary conditions) into a cone. A vertex $v_{\rm rough}$ is placed at the rough boundary (the apex of the cone), and a face $f_{\rm smooth}$ is located at the smooth boundary (the base of the cone). The following illustrates a cluster state chain of length~$4$:
\begin{equation*}
\begin{aligned}
\begin{tikzpicture}
    \def\n{3}
    \def\s{1}   
   \fill[yellow!10] (0, 0) rectangle (\n*\s+\s, \s); 

    \foreach \i in {0,...,\n} {
        \draw[-stealth, line width=1.0pt, red, midway] (\i*\s, 0) -- (\i*\s+\s, 0);
        \draw[-stealth,line width=1.0pt, midway] (\i*\s, 0) -- (\i*\s, \s);
    }
    \draw[-stealth, midway,line width=1.0pt] (\n*\s+\s, 0) -- (\n*\s+\s, \s);
    \draw[-stealth, white, line width=2.0pt, midway] (4,0) -- (4,1.02);
\end{tikzpicture}
\end{aligned} 
\Leftrightarrow
\begin{aligned}
    \begin{tikzpicture}
   
    \draw[red!60, very thick, -latex] (1.5,0) arc[start angle=0, end angle=60, x radius=1.5, y radius=0.5];

    \draw[red!60, very thick, -latex] (1.5,0) arc[start angle=0, end angle=130, x radius=1.5, y radius=0.5];
    \draw[red!60, very thick, -latex] (1.5,0) arc[start angle=0, end angle=220, x radius=1.5, y radius=0.5];
    \draw[red!60, very thick, -latex] (1.5,0) arc[start angle=0, end angle=320, x radius=1.5, y radius=0.5];
    \draw[red!60, very thick] (0,0) ellipse (1.5 and 0.5); 

    \coordinate (Apex) at (0,1.5); 


    \draw[thick, black!60] (1.52,0) -- node[midway] {} (Apex);  
    \draw[thick, black!60] (-1.52,0) -- node[midway] {} (Apex); 
    \draw[thick, black!60] (0.3,0.45) -- node[midway] {} (Apex);   
    \draw[thick, black!60] (-0.3,-0.53) -- node[midway] {} (Apex);  

    \coordinate (RightMid) at (1.52,0); 
    \draw[thick, black!60, -latex] (RightMid) -- node[midway, left] {} ($ (Apex)!0.5!(1.52,0) $);  

    \coordinate (LeftMid) at (-1.52,0); 
    \draw[thick, black!60, -latex] (LeftMid) -- node[midway, left] {} ($ (Apex)!0.5!(-1.52,0) $);  

    \coordinate (TopMid) at (0.3,0.45); 
    \draw[thick, black!60, -latex] (TopMid) -- node[midway, right] {} ($ (Apex)!0.5!(0.3,0.45) $);  

    \coordinate (BottomMid) at (-0.3,-0.53); 
    \draw[thick, black!60, -latex] (BottomMid) -- node[midway, right] {} ($ (Apex)!0.5!(-0.3,-0.53) $);  
    \node[line width=0.2pt, dashed, red, draw opacity=0.5] (a) at (0.6,0) {\(f_{\rm smooth} \)};
     \node[line width=0.2pt, dashed, draw opacity=0.5] (a) at (0,1.8) {\(v_{\rm rough} \)};
    \draw[cyan, dotted] (0,0) -- (1.52,0);  
    \draw[cyan, dotted] (0,0) -- (-1.52,0); 
    \draw[cyan, dotted] (0,0) -- (0.3,0.53);  
    \draw[cyan, dotted] (0,0) -- (-0.3,-0.53);  
    \end{tikzpicture}
    \end{aligned}
\end{equation*}
In this way, the model can be interpreted as a quantum double model on a sphere, with the top vertex operator and the bottom face operator removed.

The $\Ising$ symmetry is supported on the symmetry boundary (depicted in red). Under periodic boundary conditions, the symmetry operator associated with the smooth boundary is given by the face operator $Q_{\varphi} = \Bf^{\varphi}_{f_{\rm smooth}}$, where $\varphi \in \hTIsing$. By definition, this operator takes the form
\begin{equation}
Q_{\varphi} = \sum_{(\varphi)} \ZR_{\varphi^{(1)}} \otimes \ZR_{\varphi^{(2)}} \otimes \cdots \otimes \ZR_{\varphi^{(n)}},
\end{equation}
where $\Delta^{(n-1)}(\varphi) = \sum_{(\varphi)} \varphi^{(1)} \otimes \varphi^{(2)} \otimes \cdots \otimes \varphi^{(n)}$ is the $(n{-}1)$-fold coproduct in $\hTIsing$. More precisely, we have
\begin{equation}
\begin{aligned}
     \begin{tikzpicture}
   \fill[yellow!10] (0, 0) rectangle (5,1); 
    \draw[-stealth, line width=1.0pt, red, midway] (0,0) --(1,0);
    \draw[-stealth, line width=1.0pt, red, midway] (1,0) --(2,0);
    \draw[-stealth, line width=1.0pt, red, midway] (2,0) --(2.5,0);
    \draw[-stealth, line width=1.0pt, red, midway] (3.5,0) --(4,0);
    \draw[-stealth, line width=1.0pt, red, midway] (4,0) --(5,0);
      \draw[-stealth,line width=1.0pt, midway] (0, 0) -- (0, 1);
     \draw[-stealth,line width=1.0pt, midway] (1, 0) -- (1, 1); 
    \draw[-stealth,line width=1.0pt, midway] (2, 0) -- (2, 1); 
    \draw[-stealth,line width=1.0pt, midway] (4, 0) -- (4, 1); 
     \node[line width=0.2pt, dashed, red, draw opacity=0.5] (a) at (0.5,-0.4) {\(\ZR_{\varphi^{\cone}} \)};
    \node[line width=0.2pt, dashed, red, draw opacity=0.5] (a) at (1.5,-0.4) {\(\ZR_{\varphi^{\ctwo}} \)};
     \node[line width=0.2pt, dashed, red, draw opacity=0.5] (a) at (3,-0.4) {\(\cdots \)};
    \node[line width=0.2pt, dashed, red, draw opacity=0.5] (a) at (4.5,-0.4) {\(\ZR_{\varphi^{(n)}} \)};
    \end{tikzpicture}
\end{aligned}
\end{equation}
This gives a representation of $\hTIsing$, $Q_{\zeta} Q_{\varphi} = Q_{\zeta \cdot \varphi}$ for all $\zeta, \varphi \in \hTIsing$. When $\varphi \in \Cocom(\hTIsing)$, the operator $\Bf^{\varphi}_{f_{\rm smooth}}$ commutes with the Hamiltonian~\cite{chang2014kitaev, Jia2023weak}. This implies that the model possesses a $\Cocom(\hTIsing)$ symmetry.

Since $\Ising \simeq \Rep(\TIsing)$,  
the Grothendieck group $\operatorname{Gr}(\Ising)$  
is generated by the irreducible characters $\chi_{\Gamma}$ of $\TIsing$, and  
\begin{equation}
\chi_{\Gamma} \cdot \chi_{\Phi} = \chi_{\Gamma \otimes \Phi} = \sum_{\Psi} N_{\Gamma, \Phi}^{\Psi} \chi_{\Psi},
\end{equation}
where $N_{\Gamma, \Phi}^{\Psi} \in \mathbb{Z}_{\geq 0}$ are the fusion multiplicities.  
The character algebra is defined as $\operatorname{Char}(\Ising) = \operatorname{Gr}(\Ising) \otimes_{\mathbb{Z}} \mathbb{C}$,  
which embeds into the dual weak Hopf algebra $\hTIsing$.  
Since the character algebra lies within $\Cocom(\hTIsing)$,  
the model accordingly exhibits Ising fusion category symmetry.

The rough boundary also gives symmetry operators \( W_h = \Av_{v_{\rm rough}}^h =\sum_{(h)} \XR_{h^{\cone}} \otimes \cdots \otimes \XR_{h^{(n)}} \), with the relation \( W_h W_g = W_{hg} \), more precisely,
\begin{equation}
\begin{aligned}
     \begin{tikzpicture}
   \fill[yellow!10] (0, 0) rectangle (5,1); 
    \draw[-stealth, line width=1.0pt, red, midway] (0,0) --(1,0);
    \draw[-stealth, line width=1.0pt, red, midway] (1,0) --(2,0);
    \draw[-stealth, line width=1.0pt, red, midway] (2,0) --(2.5,0);
    \draw[-stealth, line width=1.0pt, red, midway] (3.5,0) --(4,0);
    \draw[-stealth, line width=1.0pt, red, midway] (4,0) --(5,0);
      \draw[-stealth,line width=1.0pt, midway] (0, 0) -- (0, 1);
     \draw[-stealth,line width=1.0pt, midway] (1, 0) -- (1, 1); 
    \draw[-stealth,line width=1.0pt, midway] (2, 0) -- (2, 1); 
    \draw[-stealth,line width=1.0pt, midway] (4, 0) -- (4, 1); 
     \node[line width=0.2pt, dashed, draw opacity=0.5] (a) at (0.5,0.5) {\(\XR_{h^{\cone}} \)};
    \node[line width=0.2pt, dashed, draw opacity=0.5] (a) at (1.5,0.5) {\(\XR_{h^{\ctwo}} \)};
    \node[line width=0.2pt, dashed, draw opacity=0.5] (a) at (2.5,0.5) {\(\XR_{h^{\cthree}} \)};
     \node[line width=0.2pt, dashed, draw opacity=0.5] (a) at (3.4,0.5) {\(\cdots \)};
    \node[line width=0.2pt, dashed, draw opacity=0.5] (a) at (4.5,0.5) {\(\XR_{h^{(n)}} \)};
    \end{tikzpicture}
\end{aligned}
\end{equation}
When \( h \in \Cocom(\TIsing) \), \( W_h \) commutes with the Hamiltonian. This implies that the model has a \( \Cocom(\TIsing) \) symmetry.

\subsection{Ribbon operators and edge modes}

For Ising cluster state model,  the string order parameters are given by ribbon operators. If we apply ribbon operators on the cluster state, there will be two excitations on the boundary of ribbon created by ribbon operators.
\begin{equation*}
   \begin{aligned}
\begin{tikzpicture}
    \def\n{5}
    \def\s{1}
    \fill[yellow!20] (-1, 0) rectangle (7,1); 
    \foreach \i in {0,...,\n} {
        \draw[-stealth, line width=1.0pt,red, midway] (\i*\s, 0) -- (\i*\s+\s, 0);
        \draw[dotted, line width=1.0pt,white, midway] (\i*\s, \s) -- (\i*\s+\s, \s);
        \draw[-stealth,line width=1.0pt, midway] (\i*\s, 0) -- (\i*\s, \s);
    }
    \draw[-stealth,line width=1.0pt, midway] (-1, 0) -- (-1, 1);
    \draw[-stealth,line width=1.0pt, midway] (7, 0) -- (7, 1);
    \draw[-stealth, midway,line width=1.0pt] (\n*\s+\s, 0) -- (\n*\s+\s, \s);
  \draw[-stealth, line width=1.0pt, red, midway] (-1,0) --(0,0);
  \draw[-stealth, line width=1.0pt, red, midway] (6,0) --(7,0);
\filldraw[purple!10, opacity=0.7] (0,0) -- (-0.5,0.5) -- (6.5, 0.5) -- (6,0) -- cycle;

   \draw[black,dashed, line width = 0.7pt] (-0.5,0.5) -- (6.5,0.5);
   \draw[cyan,dotted, line width = 1pt] (0,0) -- (-0.5,0.5);
   \draw[cyan,dotted, line width = 1pt] (0,0) -- (0.5,0.5);
  \draw[cyan,dotted, line width = 1pt] (1,0) -- (0.5,0.5);
  \draw[cyan,dotted, line width = 1pt] (1,0) -- (1.5,0.5);
    \draw[cyan,dotted, line width = 1pt] (2,0) -- (1.5,0.5);
        \draw[cyan,dotted, line width = 1pt] (2,0) -- (2.5,0.5);
          \draw[cyan,dotted, line width = 1pt] (3,0) -- (2.5,0.5);
                \draw[cyan,dotted, line width = 1pt] (3,0) -- (3.5,0.5);
 \draw[cyan,dotted, line width = 1pt] (4,0) -- (3.5,0.5);
\draw[cyan,dotted, line width = 1pt] (4,0) -- (4.5,0.5);
\draw[cyan,dotted, line width = 1pt] (5,0) -- (4.5,0.5);
\draw[cyan,dotted, line width = 1pt] (5,0) -- (5.5,0.5);
\draw[cyan,dotted, line width = 1pt] (6,0) -- (5.5,0.5);
\draw[cyan,dotted, line width = 1pt] (6,0) -- (6.5,0.5);
\end{tikzpicture}
\end{aligned} 
\end{equation*}
The black dashed line represent the dual lattice edges, its orientation is obtained by rotating the direct edge with $\pi/2$, and cyan dotted line represents the sites $s=(v_s,f)$.

A direct triangle $\tau = (s_0, s_1, e)$ consists of two sites $s_0, s_1$ and a direct edge $e$. We assign a direction to $\tau$ such that its starting and terminal sites are $\partial_0 \tau = s_0$ and $\partial_1 \tau = s_1$, respectively. Similarly, a dual triangle is denoted by $\tilde{\tau} = (s_0, s_1, \tilde{e})$. 
A ribbon $\rho = (\rho_1, \ldots, \rho_n)$ is a sequence of triangles $\rho_i$, each of which may be either direct or dual, satisfying $\partial_1 \rho_i = \partial_0 \rho_{i+1}$ and having no self-overlap.

The ribbon operator is defined recursively, starting from the triangle operator and extended through a recursive relation. For a triangle, the operator is given by (for $h, x \in \TIsing$ and $\varphi \in \hTIsing$)
\begin{align}
	&	\begin{aligned}
		\begin{tikzpicture}
			\draw[-latex,black] (1,0) -- (-1,0); 
			\node[ line width=0.2pt, dashed, draw opacity=0.5] (a) at (0,0.2){$x$};
			\draw[dotted, cyan, line width=0.5pt] (-1,0) -- (0,-1);
			\node[ line width=0.2pt, dashed, draw opacity=0.5] (a) at (-0.6,-0.7){$s_1$};
			\draw[dotted, cyan, line width=0.5pt] (1,0) -- (0,-1);
			\node[ line width=0.2pt, dashed, draw opacity=0.5] (a) at (0.6,-0.7){$s_0$};
			\draw[-stealth,gray, line width=3pt] (0.5,-0.4) -- (-0.5,-0.4); 
		\end{tikzpicture}
	\end{aligned}
	\quad
	\begin{aligned}
		& F^{h,\varphi}( \tau_R)|x\rangle 
        =  
		\varepsilon(h)  \ZL_{\varphi} |x\rangle \\
		= & \varepsilon(h) | x \leftharpoonup \hat{S}( \varphi) \rangle,
	\end{aligned} \label{eq:tri1} \\
	&\begin{aligned}
		\begin{tikzpicture}
			\draw[-latex,black] (-1,0) -- (1,0); 
			\node[ line width=0.2pt, dashed, draw opacity=0.5] (a) at (0,0.2){$x$};
			\draw[dotted, cyan, line width=0.5pt] (-1,0) -- (0,-1);
			\node[ line width=0.2pt, dashed, draw opacity=0.5] (a) at (-0.6,-0.7){$s_1$};
			\draw[dotted, cyan, line width=0.5pt] (1,0) -- (0,-1);
			\node[ line width=0.2pt, dashed, draw opacity=0.5] (a) at (0.6,-0.7){$s_0$};
			\draw[-stealth,gray, line width=3pt] (0.5,-0.4) -- (-0.5,-0.4); 
		\end{tikzpicture}
	\end{aligned}
	\quad
	\begin{aligned}
		& F^{h,\varphi}( \tau_R)|x\rangle 
        =  	\varepsilon(h)  \ZR_{\varphi} |x\rangle\\
	&=\varepsilon(h) | \varphi \rightharpoonup x \rangle.
	\end{aligned}
\end{align}
	\begin{align} 
		&	\begin{aligned}
			\begin{tikzpicture}
				\draw[-latex,dashed,black] (1,0) -- (-1,0); 
				\node[ line width=0.2pt, dashed, draw opacity=0.5] (a) at (0,0.2){$x$};
				\draw[dotted, cyan, line width=0.5pt] (-1,0) -- (0,-1);
				\node[ line width=0.2pt, dashed, draw opacity=0.5] (a) at (-0.6,-0.7){$s_1$};
				\draw[dotted, cyan, line width=0.5pt] (1,0) -- (0,-1);
				\node[ line width=0.2pt, dashed, draw opacity=0.5] (a) at (0.6,-0.7){$s_0$};
				\draw[-stealth,gray, line width=3pt] (0.5,-0.4) -- (-0.5,-0.4); 
			\end{tikzpicture}
		\end{aligned}
		\quad
		\begin{aligned}
			& F^{h,\varphi}( \tilde{\tau}_R)|x\rangle =   \hat{\varepsilon}(\varphi) \XL_h |x\rangle \\
			& =  \hat{\varepsilon}(\varphi) |  x \triangleleft  S(h) \rangle  ,
		\end{aligned}\\
		&\begin{aligned}
			\begin{tikzpicture}
				\draw[-latex,dashed,black] (-1,0) -- (1,0); 
				\node[ line width=0.2pt, dashed, draw opacity=0.5] (a) at (0,0.2){$x$};
				\draw[dotted, cyan, line width=0.5pt] (-1,0) -- (0,-1);
				\node[ line width=0.2pt, dashed, draw opacity=0.5] (a) at (-0.6,-0.7){$s_1$};
				\draw[dotted, cyan, line width=0.5pt] (1,0) -- (0,-1);
				\node[ line width=0.2pt, dashed, draw opacity=0.5] (a) at (0.6,-0.7){$s_0$};
				\draw[-stealth,gray, line width=3pt] (0.5,-0.4) -- (-0.5,-0.4); 
			\end{tikzpicture}
		\end{aligned}
		\quad
		\begin{aligned}
		&	F^{h,\varphi}( \tilde{\tau}_R)|x\rangle =  
		\hat{\varepsilon}(\varphi) \XR_h |x\rangle\\
            & = \hat{\varepsilon}(\varphi) | h\triangleright x \rangle.
		\end{aligned} \label{eq:tri8}
	\end{align}
\begin{align}
		&	\begin{aligned}
			\begin{tikzpicture}
				\draw[-latex,dashed,black] (1,0) -- (-1,0); 
				\node[ line width=0.2pt, dashed, draw opacity=0.5] (a) at (0,0.2){$x$};
				\draw[dotted, cyan, line width=0.5pt] (-1,0) -- (0,-1);
				\node[ line width=0.2pt, dashed, draw opacity=0.5] (a) at (-0.6,-0.7){$s_0$};
				\draw[dotted, cyan, line width=0.5pt] (1,0) -- (0,-1);
				\node[ line width=0.2pt, dashed, draw opacity=0.5] (a) at (0.6,-0.7){$s_1$};
				\draw[-stealth,gray, line width=3pt] (-0.5,-0.4) -- (0.5,-0.4); 
			\end{tikzpicture}
		\end{aligned}
		\quad
		\begin{aligned}
		&	F^{h,\varphi}( \tilde{\tau}_L)|x\rangle =  \hat{\varepsilon}(\varphi)   \tilde{\XL}_h |x\rangle \\
     &   = \hat{\varepsilon}(\varphi)  
			| x\triangleleft  h \rangle.
		\end{aligned}\\
		&\begin{aligned}
			\begin{tikzpicture}
				\draw[-latex,dashed,black] (-1,0) -- (1,0); 
				\node[ line width=0.2pt, dashed, draw opacity=0.5] (a) at (0,0.2){$x$};
				\draw[dotted, cyan, line width=0.5pt] (-1,0) -- (0,-1);
				\node[ line width=0.2pt, dashed, draw opacity=0.5] (a) at (-0.6,-0.7){$s_0$};
				\draw[dotted, cyan, line width=0.5pt] (1,0) -- (0,-1);
				\node[ line width=0.2pt, dashed, draw opacity=0.5] (a) at (0.6,-0.7){$s_1$};
				\draw[-stealth,gray, line width=3pt] (-0.5,-0.4) -- (0.5,-0.4); 
			\end{tikzpicture}
		\end{aligned}
		\quad
		\begin{aligned}
		&	F^{h,\varphi}( \tilde{\tau}_L)|x\rangle =   \hat{\varepsilon}(\varphi) \tilde{\XR}_h |x\rangle \\
		&	= \hat{\varepsilon}(\varphi)    	
			|S(h)\triangleright  x\rangle.
		\end{aligned}
	\end{align}
	\begin{align}
		&	\begin{aligned}
			\begin{tikzpicture}
				\draw[-latex,black] (1,0) -- (-1,0); 
				\node[ line width=0.2pt, dashed, draw opacity=0.5] (a) at (0,0.2){$x$};
				\draw[dotted, cyan, line width=0.5pt] (-1,0) -- (0,-1);
				\node[ line width=0.2pt, dashed, draw opacity=0.5] (a) at (-0.6,-0.7){$s_0$};
				\draw[dotted, cyan, line width=0.5pt] (1,0) -- (0,-1);
				\node[ line width=0.2pt, dashed, draw opacity=0.5] (a) at (0.6,-0.7){$s_1$};
				\draw[-stealth,gray, line width=3pt] (-0.5,-0.4) -- (0.5,-0.4); 
			\end{tikzpicture}
		\end{aligned}
		\quad
		\begin{aligned}
		&	F^{h,\varphi}( \tau_L)|x\rangle =  
			\varepsilon(h)  \tilde{\ZL}_{\varphi} |x\rangle\\
			& =  \varepsilon(h) 
			| x \leftharpoonup \varphi \rangle ,
		\end{aligned}\\
		&\begin{aligned}
			\begin{tikzpicture}
				\draw[-latex,black] (-1,0) -- (1,0); 
				\node[ line width=0.2pt, dashed, draw opacity=0.5] (a) at (0,0.2){$x$};
				\draw[dotted, cyan, line width=0.5pt] (-1,0) -- (0,-1);
				\node[ line width=0.2pt, dashed, draw opacity=0.5] (a) at (-0.6,-0.7){$s_0$};
				\draw[dotted, cyan, line width=0.5pt] (1,0) -- (0,-1);
				\node[ line width=0.2pt, dashed, draw opacity=0.5] (a) at (0.6,-0.7){$s_1$};
				\draw[-stealth,gray, line width=3pt] (-0.5,-0.4) -- (0.5,-0.4); 
			\end{tikzpicture}
		\end{aligned}
		\quad
		\begin{aligned}
			&F^{h,\varphi}( \tau_L)|x\rangle = 
				\varepsilon(h)  \tilde{\ZR}_{\varphi} |x\rangle \\
		&	= \varepsilon(h) 	|  \hat{S}( \varphi) \rightharpoonup  x \rangle.
		\end{aligned}  \label{eq:tri16}
	\end{align}

Consider an element $h \otimes \varphi \in \TIsing \otimes \hTIsing$ and a ribbon $\rho$. The ribbon operator $F^{h,\varphi}(\rho)$ acts non-trivially only on the edges along $\rho$.
Suppose the ribbon decomposes as $\rho = \rho_1 \cup \rho_2$, where both $\rho_1$ and $\rho_2$ share the orientation of $\rho$, and satisfy $\partial_1 \rho_1 = \partial_0 \rho_2$. Then, the ribbon operator on $\rho$ is defined by
\begin{equation}\begin{aligned}
  &  F^{h,\varphi}(\rho) = \sum_{(h \otimes \varphi)} F^{(h \otimes \varphi)^{(1)}}(\rho_1) F^{(h \otimes \varphi)^{(2)}}(\rho_2) \\
= &\sum_k \sum_{(k), (h)} F^{h^{(1)}, \hat{k}}(\rho_1) F^{S(k^{(3)}) h^{(2)} k^{(1)}, \varphi(k^{(2)} \bullet)}(\rho_2).
\end{aligned}
\end{equation}
Here, $\{k\}$ is an orthogonal basis of $\TIsing$ with dual basis $\{\hat{k}\}$. The first equality uses the comultiplication in the dual weak Hopf algebra $D(\TIsing)^{\vee} \cong \mathbf{Tube}({}_{\Ising}\Ising_{\Ising})^{\vee}$, which implies the second equality.
This definition does not depend on the particular decomposition $\rho = \rho_1 \cup \rho_2$. After iterating this decomposition enough times, the ribbon breaks down into triangles, and the ribbon operator is fully determined by the previously defined triangle operators.

Let $\rho$ be a ribbon. If $h \in \operatorname{Cocom}(\TIsing)$ and $\varphi \in \operatorname{Cocom}(\hTIsing)$, then for all sites $s \neq \partial_0 \rho, \partial_1 \rho$, the operators $\Av^h(s)$ and $\Bf^\varphi(s)$ commute with $F^{g,\psi}(\rho)$:
\begin{equation}
\begin{aligned}
    &\Av^{\lambda}(s) F^{g,\psi}(\rho) = F^{g,\psi}(\rho) \Av^{\lambda}(s), \\
    &\Bf^{\Lambda}(s) F^{g,\psi}(\rho) = F^{g,\psi}(\rho) \Bf^{\Lambda}(s).
\end{aligned}
\end{equation}
However, at the boundary sites $s = (v_s,f)$ of $\rho$, the commutators
\begin{equation}
[F^{g,\psi}(\rho), \Av_{v_s}] \neq 0, \quad [F^{g,\psi}(\rho), \Bf_f] \neq 0,
\end{equation}
show that the ribbon operators create excitations localized at these boundary sites.

We can also place the cluster state model on an open spatial manifold (equivalently, make a cut of the cluster state on $\mathbb{S}^1$). In this case, edge modes will appear.
Note that on an open manifold, the symmetry strongly depends on the choice of cut made to the closed-manifold cluster state model \cite{jia2024weakhopfnoninvertible}.
One special cut is as follows
\begin{equation*}
   \begin{aligned}
\begin{tikzpicture}
    \def\n{5}
    \def\s{1}
    \fill[yellow!20] (-1, 0) rectangle (7,1); 
    \foreach \i in {0,...,\n} {
        \draw[-stealth, line width=1.0pt,red, midway] (\i*\s, 0) -- (\i*\s+\s, 0);
        \draw[dotted, line width=1.0pt,white, midway] (\i*\s, \s) -- (\i*\s+\s, \s);
        \draw[-stealth,line width=1.0pt, midway] (\i*\s, 0) -- (\i*\s, \s);
    }
    \draw[-stealth, midway,line width=1.0pt] (\n*\s+\s, 0) -- (\n*\s+\s, \s);
  \draw[-stealth, line width=1.0pt, red, midway] (-1,0) --(0,0);
  \draw[-stealth, line width=1.0pt, red, midway] (6,0) --(7,0);
\filldraw[purple!10, opacity=0.7] (-1,0) -- (-0.5,0.5) -- (6.5, 0.5) -- (7,0) -- cycle;

   \draw[black,dashed, line width = 0.7pt] (-0.5,0.5) -- (6.5,0.5);
   \draw[cyan,dotted, line width = 1pt] (-1,0) -- (-0.5,0.5);
   \draw[cyan,dotted, line width = 1pt] (0,0) -- (-0.5,0.5);
   \draw[cyan,dotted, line width = 1pt] (0,0) -- (0.5,0.5);
  \draw[cyan,dotted, line width = 1pt] (1,0) -- (0.5,0.5);
  \draw[cyan,dotted, line width = 1pt] (1,0) -- (1.5,0.5);
    \draw[cyan,dotted, line width = 1pt] (2,0) -- (1.5,0.5);
        \draw[cyan,dotted, line width = 1pt] (2,0) -- (2.5,0.5);
          \draw[cyan,dotted, line width = 1pt] (3,0) -- (2.5,0.5);
                \draw[cyan,dotted, line width = 1pt] (3,0) -- (3.5,0.5);
 \draw[cyan,dotted, line width = 1pt] (4,0) -- (3.5,0.5);
\draw[cyan,dotted, line width = 1pt] (4,0) -- (4.5,0.5);
\draw[cyan,dotted, line width = 1pt] (5,0) -- (4.5,0.5);
\draw[cyan,dotted, line width = 1pt] (5,0) -- (5.5,0.5);
\draw[cyan,dotted, line width = 1pt] (6,0) -- (5.5,0.5);
\draw[cyan,dotted, line width = 1pt] (6,0) -- (6.5,0.5);
\draw[cyan,dotted, line width = 1pt] (7,0) -- (6.5,0.5);
\end{tikzpicture}
\end{aligned} 
\end{equation*}
In this case, the symmetry is enhanced because the ribbon operators $F^{g,\varphi}(\rho)$ fail to commute only with the vertex and face operators at the boundary sites. However, since both ends of the cut no longer have such operators, the model realizes the full quantum double symmetry $D(\TIsing)^{\vee} \cong \mathbf{Tube}({}_{\Ising}\Ising_{\Ising})^{\vee}$.
Consequently, $\TIsing $ and $\hTIsing$ appear as sub-symmetries in this scenario. Since $\operatorname{Gr}(\Ising) \hookrightarrow \hTIsing$, the model also have $\Ising$ symmetry.


\section{Weak Hopf tensor network for Ising cluster state model}
\label{sec:IsingCluStateTN}

The ground state of the Ising cluster state model will be referred to as the Ising cluster state.  
This cluster state model can be solved using the weak Hopf matrix product state formalism developed in Refs.~\cite{jia2024generalized,jia2024weakhopfnoninvertible,Jia2023weak,jia2023boundary,Buerschaper2013a}.  
To this end, we first introduce some basic local tensors.

For edges on the smooth boundary, the local tensor is chosen as the comultiplication of the Haar integral $\lambda \in \TIsing$:
\begin{equation}
     \Delta(\lambda) = 
     \begin{aligned}
\begin{tikzpicture}
    \node[draw, fill=blue!20, minimum width=0.6cm, minimum height=0.6cm] (center) at (0,0) {$\scriptstyle\lambda$};
    \draw[line width=1.0pt] (center.north) -- ++(0,.5) node[above] {$\scriptstyle\lambda^{\ctwo}$};
    \draw[line width=1.0pt] (center.south) -- ++(0,-.5) node[below] {$\scriptstyle \lambda^{\cone}$};
\end{tikzpicture}
\end{aligned}.
\end{equation}
The local tensor for a bulk edge corresponds to $\Delta_2(\lambda)$:
\begin{equation}
     \Delta(\lambda)= \begin{aligned}			
\begin{tikzpicture}
    \draw[line width=1.0pt] (center.north) (0,0) -- ++(0.8,0) node[right] {$\scriptstyle\lambda^{\ctwo}$};
    \draw[line width=1.0pt] (center.north) (0,0) -- ++(-0.8,0) node[left] {$\scriptstyle\lambda^{\cthree}$};
    \draw[line width=1.0pt ] (center.south) -- ++(0,-.5) node[below] {$\scriptstyle \lambda^{\cone}$};
    \node[draw, fill=gray!20, minimum width=0.6cm, minimum height=0.6cm] (center) at (0,0) {$\scriptstyle\lambda$};
\end{tikzpicture}
\end{aligned}.
\end{equation}
Since $\lambda$ is cocommutative, the labels of its components can be permuted cyclically.

The local tensor for a face corresponds to the Haar measure $\Lambda \in \hTIsing$, which is introduced to glue the edge tensors together:
\begin{equation}
    ( \id \otimes \id\otimes \hat{S}) \circ \hat{\Delta}(\Lambda)=
    \begin{aligned}			
\begin{tikzpicture}
\draw[line width=1.0pt] (center.north) (0,0) -- ++(0.8,0) node[right] {$\scriptstyle\Lambda^{\ctwo}$};
\draw[line width=1.0pt] (center.north) (0,0) -- ++(-0.8,0) node[left] {$\scriptstyle \hat{S}(\Lambda^{\cthree})$};
\draw[line width=1.0pt ] (center.south) -- ++(0,-.5) node[below] {$\scriptstyle \Lambda^{\cone}$};
\node[draw, fill=yellow!20, minimum size=0.6cm, shape=circle] (center) at (0,0) {$\scriptstyle\Lambda$};
 \filldraw[black] (-0.5,0) circle (2pt);  
\end{tikzpicture}
\end{aligned}. 
\end{equation}
Here we use black dot to represent taking antipode.

Since the face tensor in in the dual algebra, thus gluing the face tensor and edge tensor means pairing.
The Ising cluster state is defined as the following tensor network state:
\begin{equation}
    \begin{aligned}			
\begin{tikzpicture}
    \draw[line width=1.0pt] (-0.5,0) -- ++(7.3,0) ;
    \draw[line width=1.0pt, red] (0,0) -- ++(0,-.8);
        \draw[line width=1.0pt, red] (2,0) -- ++(0,-.8);
    \draw[line width=1.0pt, red] (4,0) -- ++(0,-.8);
    \draw[line width=1.0pt, red] (6,0) -- ++(0,-.8);
     \filldraw[black] (0.5,0) circle (2pt);  
      \filldraw[black] (2.5,0) circle (2pt);
       \filldraw[black] (4.5,0) circle (2pt);
        \filldraw[black] (6.5,0) circle (2pt);
    \draw[line width=1.0pt] (1,-1) -- ++(0,1) ;
    \draw[line width=1.0pt, red] (1,-1) -- ++(0,-.8);
    \node[draw, fill=blue!20, minimum width=0.6cm, minimum height=0.6cm] (center) at (1,-1) {$\scriptstyle\lambda$};
     \draw[line width=1.0pt] (3,-1) -- ++(0,1) ;
    \draw[line width=1.0pt, red] (3,-1) -- ++(0,-.8);
    \node[draw, fill=blue!20, minimum width=0.6cm, minimum height=0.6cm] (center) at (3,-1) {$\scriptstyle\lambda$};
         \draw[line width=1.0pt] (5,-1) -- ++(0,1) ;
    \draw[line width=1.0pt, red] (5,-1) -- ++(0,-.8);
    \node[draw, fill=blue!20, minimum width=0.6cm, minimum height=0.6cm] (center) at (5,-1) {$\scriptstyle\lambda$};
    \node[draw, fill=gray!20, minimum width=0.6cm, minimum height=0.6cm] (center) at (0,0) {$\scriptstyle\lambda$};
    \node[draw, fill=yellow!20, minimum size=0.6cm, shape=circle] (center) at (1,0) {$\scriptstyle\Lambda$};
     \node[draw, fill=gray!20, minimum width=0.6cm, minimum height=0.6cm] (center) at (2,0) {$\scriptstyle\lambda$};
      \node[draw, fill=yellow!20, minimum size=0.6cm, shape=circle] (center) at (3,0) {$\scriptstyle\Lambda$};
        \node[draw, fill=gray!20, minimum width=0.6cm, minimum height=0.6cm] (center) at (4,0) {$\scriptstyle\lambda$};
      \node[draw, fill=yellow!20, minimum size=0.6cm, shape=circle] (center) at (5,0) {$\scriptstyle\Lambda$};
        \node[draw, fill=gray!20, minimum width=0.6cm, minimum height=0.6cm] (center) at (6,0) {$\scriptstyle\lambda$};
\end{tikzpicture}
\end{aligned} 
\end{equation}
The red free leg represents the physical degrees of freedom. This tensor network forms the ground state of the Ising cluster state model; a general proof is provided in our previous work~\cite{Jia2023weak}.

\section{Discussion and conclusion}

This note provides a concrete example to illustrate our method in Ref.~\cite{jia2024generalized} for constructing lattice models with non-invertible symmetries. We present a construction of a cluster state model with $\Ising$ symmetry, based on the Ising boundary tube algebra. Since the model admits an explicit basis representation, numerical investigation of its properties becomes feasible—although computationally demanding, given the local Hilbert space dimension is $34$.

From a SymTFT perspective, we think that the Ising cluster state model constructed here does not realize a symmetry-protected topological (SPT) phase, but rather a (partial) spontaneous symmetry-breaking (SSB) phase. The ground state may not be unique, as the rough boundary corresponds to a simple removal of boundary degrees of freedom. However, a detailed analysis is still needed to determine the precise ground state degeneracy.

Understanding how non-invertible symmetry anomalies manifest on the lattice remains a significant open question. A crucial first step in this direction is to clarify the precise meaning of the “on-site” nature of such symmetries. Moreover, it is important to distinguish between fusion algebra symmetry and fusion category symmetry at the lattice level, as they may exhibit fundamentally different behaviors in microscopic realizations.

We leave these problems to future work.

\subsection*{Acknowledgements}

This paper arises from a discussion with Salvatore Pace during the ``TSVP Symposium: Aspects of Generalized Symmetries'' at the Okinawa Institute of Science and Technology.
The author also thanks Daniel Bulmash, Hiromi Ebisu, Luisa Eck, Jung Hoon Han, Sheng-Jie Huang, Kansei Inamura, Qiang Jia, Ryohei Kobayashi, Liang Kong, Yingcheng Li, Da-Chuan Lu, Ran Luo, Salvatore Pace,  Marvin Qi, Ziyun Xu, Masahito Yamazaki, Xinping Yang, Peng Ye, Liujun Zou for beneficial communications on SymTFT during his stay at OIST and elsewhere.
The author also acknowledges Dagomir Kaszlikowski for his support. This work is supported by the National Research Foundation in Singapore, the A*STAR under its CQT Bridging Grant, CQT-Return of PIs EOM YR1-10 Funding and  and  CQT Young Researcher Career Development Grant.


\bibliographystyle{apsrev4-1-title}
\bibliography{jiabib.bib}

\begin{thebibliography}{72}%
\makeatletter
\providecommand \@ifxundefined [1]{%
 \@ifx{#1\undefined}
}%
\providecommand \@ifnum [1]{%
 \ifnum #1\expandafter \@firstoftwo
 \else \expandafter \@secondoftwo
 \fi
}%
\providecommand \@ifx [1]{%
 \ifx #1\expandafter \@firstoftwo
 \else \expandafter \@secondoftwo
 \fi
}%
\providecommand \natexlab [1]{#1}%
\providecommand \enquote  [1]{``#1''}%
\providecommand \bibnamefont  [1]{#1}%
\providecommand \bibfnamefont [1]{#1}%
\providecommand \citenamefont [1]{#1}%
\providecommand \href@noop [0]{\@secondoftwo}%
\providecommand \href [0]{\begingroup \@sanitize@url \@href}%
\providecommand \@href[1]{\@@startlink{#1}\@@href}%
\providecommand \@@href[1]{\endgroup#1\@@endlink}%
\providecommand \@sanitize@url [0]{\catcode `\\12\catcode `\$12\catcode
  `\&12\catcode `\#12\catcode `\^12\catcode `\_12\catcode `\%12\relax}%
\providecommand \@@startlink[1]{}%
\providecommand \@@endlink[0]{}%
\providecommand \url  [0]{\begingroup\@sanitize@url \@url }%
\providecommand \@url [1]{\endgroup\@href {#1}{\urlprefix }}%
\providecommand \urlprefix  [0]{URL }%
\providecommand \Eprint [0]{\href }%
\providecommand \doibase [0]{http://dx.doi.org/}%
\providecommand \selectlanguage [0]{\@gobble}%
\providecommand \bibinfo  [0]{\@secondoftwo}%
\providecommand \bibfield  [0]{\@secondoftwo}%
\providecommand \translation [1]{[#1]}%
\providecommand \BibitemOpen [0]{}%
\providecommand \bibitemStop [0]{}%
\providecommand \bibitemNoStop [0]{.\EOS\space}%
\providecommand \EOS [0]{\spacefactor3000\relax}%
\providecommand \BibitemShut  [1]{\csname bibitem#1\endcsname}%
\let\auto@bib@innerbib\@empty
\bibitem [{\citenamefont {Cordova}\ \emph {et~al.}(2022)\citenamefont
  {Cordova}, \citenamefont {Dumitrescu}, \citenamefont {Intriligator},\ and\
  \citenamefont {Shao}}]{cordova2022snowmass}%
  \BibitemOpen
  \bibfield  {author} {\bibinfo {author} {\bibfnamefont {C.}~\bibnamefont
  {Cordova}}, \bibinfo {author} {\bibfnamefont {T.~T.}\ \bibnamefont
  {Dumitrescu}}, \bibinfo {author} {\bibfnamefont {K.}~\bibnamefont
  {Intriligator}}, \ and\ \bibinfo {author} {\bibfnamefont {S.-H.}\
  \bibnamefont {Shao}},\ }\href@noop {} {\enquote {\bibinfo {title} {Snowmass
  white paper: Generalized symmetries in quantum field theory and beyond},}\ }
  (\bibinfo {year} {2022}),\ \Eprint {http://arxiv.org/abs/2205.09545}
  {arXiv:2205.09545 [hep-th]} \BibitemShut {NoStop}%
\bibitem [{\citenamefont {Brennan}\ and\ \citenamefont
  {Hong}(2023)}]{brennan2023introduction}%
  \BibitemOpen
  \bibfield  {author} {\bibinfo {author} {\bibfnamefont {T.~D.}\ \bibnamefont
  {Brennan}}\ and\ \bibinfo {author} {\bibfnamefont {S.}~\bibnamefont {Hong}},\
  }\href@noop {} {\enquote {\bibinfo {title} {Introduction to generalized
  global symmetries in {QFT} and particle physics},}\ } (\bibinfo {year}
  {2023}),\ \Eprint {http://arxiv.org/abs/2306.00912} {arXiv:2306.00912
  [hep-ph]} \BibitemShut {NoStop}%
\bibitem [{\citenamefont {McGreevy}(2023)}]{mcgreevy2023generalized}%
  \BibitemOpen
  \bibfield  {author} {\bibinfo {author} {\bibfnamefont {J.}~\bibnamefont
  {McGreevy}},\ }\bibfield  {title} {\enquote {\bibinfo {title} {Generalized
  symmetries in condensed matter},}\ }\href
  {https://www.annualreviews.org/content/journals/10.1146/annurev-conmatphys-040721-021029}
  {\bibfield  {journal} {\bibinfo  {journal} {Annual Review of Condensed Matter
  Physics}\ }\textbf {\bibinfo {volume} {14}},\ \bibinfo {pages} {57} (\bibinfo
  {year} {2023})},\ \Eprint {http://arxiv.org/abs/2204.03045} {arXiv:2204.03045
  [cond-mat.str-el]} \BibitemShut {NoStop}%
\bibitem [{\citenamefont {Luo}\ \emph {et~al.}(2024)\citenamefont {Luo},
  \citenamefont {Wang},\ and\ \citenamefont {Wang}}]{luo2023lecture}%
  \BibitemOpen
  \bibfield  {author} {\bibinfo {author} {\bibfnamefont {R.}~\bibnamefont
  {Luo}}, \bibinfo {author} {\bibfnamefont {Q.-R.}\ \bibnamefont {Wang}}, \
  and\ \bibinfo {author} {\bibfnamefont {Y.-N.}\ \bibnamefont {Wang}},\
  }\bibfield  {title} {\enquote {\bibinfo {title} {Lecture notes on generalized
  symmetries and applications},}\ }\href {\doibase
  https://doi.org/10.1016/j.physrep.2024.02.002} {\bibfield  {journal}
  {\bibinfo  {journal} {Physics Reports}\ }\textbf {\bibinfo {volume} {1065}},\
  \bibinfo {pages} {1} (\bibinfo {year} {2024})},\ \Eprint
  {http://arxiv.org/abs/2307.09215} {arXiv:2307.09215 [hep-th]} \BibitemShut
  {NoStop}%
\bibitem [{\citenamefont {Gomes}(2023)}]{gomes2023introduction}%
  \BibitemOpen
  \bibfield  {author} {\bibinfo {author} {\bibfnamefont {P.~R.}\ \bibnamefont
  {Gomes}},\ }\bibfield  {title} {\enquote {\bibinfo {title} {An introduction
  to higher-form symmetries},}\ }\href
  {https://scipost.org/10.21468/SciPostPhysLectNotes.74} {\bibfield  {journal}
  {\bibinfo  {journal} {SciPost Physics Lecture Notes}\ ,\ \bibinfo {pages}
  {074}} (\bibinfo {year} {2023})},\ \Eprint {http://arxiv.org/abs/2303.01817}
  {arXiv:2303.01817 [hep-th]} \BibitemShut {NoStop}%
\bibitem [{\citenamefont {Shao}(2024)}]{shao2024whats}%
  \BibitemOpen
  \bibfield  {author} {\bibinfo {author} {\bibfnamefont {S.-H.}\ \bibnamefont
  {Shao}},\ }\href@noop {} {\enquote {\bibinfo {title} {What's done cannot be
  undone: Tasi lectures on non-invertible symmetries},}\ } (\bibinfo {year}
  {2024}),\ \Eprint {http://arxiv.org/abs/2308.00747} {arXiv:2308.00747
  [hep-th]} \BibitemShut {NoStop}%
\bibitem [{\citenamefont {Sch\"{a}fer-Nameki}(2024)}]{SchaferNameki2024ICTP}%
  \BibitemOpen
  \bibfield  {author} {\bibinfo {author} {\bibfnamefont {S.}~\bibnamefont
  {Sch\"{a}fer-Nameki}},\ }\bibfield  {title} {\enquote {\bibinfo {title}
  {{ICTP} lecture on (non-)invertible generalized symmetries},}\ }\href
  {\doibase https://doi.org/10.1016/j.physrep.2024.01.007} {\bibfield
  {journal} {\bibinfo  {journal} {Physics Reports}\ }\textbf {\bibinfo {volume}
  {1063}},\ \bibinfo {pages} {1} (\bibinfo {year} {2024})},\ \Eprint
  {http://arxiv.org/abs/2305.18296} {arXiv:2305.18296 [hep-th]} \BibitemShut
  {NoStop}%
\bibitem [{\citenamefont {Bhardwaj}\ \emph
  {et~al.}(2024{\natexlab{a}})\citenamefont {Bhardwaj}, \citenamefont
  {Bottini}, \citenamefont {Fraser-Taliente}, \citenamefont {Gladden},
  \citenamefont {Gould}, \citenamefont {Platschorre},\ and\ \citenamefont
  {Tillim}}]{Bhardwaj2024lecture}%
  \BibitemOpen
  \bibfield  {author} {\bibinfo {author} {\bibfnamefont {L.}~\bibnamefont
  {Bhardwaj}}, \bibinfo {author} {\bibfnamefont {L.~E.}\ \bibnamefont
  {Bottini}}, \bibinfo {author} {\bibfnamefont {L.}~\bibnamefont
  {Fraser-Taliente}}, \bibinfo {author} {\bibfnamefont {L.}~\bibnamefont
  {Gladden}}, \bibinfo {author} {\bibfnamefont {D.~S.}\ \bibnamefont {Gould}},
  \bibinfo {author} {\bibfnamefont {A.}~\bibnamefont {Platschorre}}, \ and\
  \bibinfo {author} {\bibfnamefont {H.}~\bibnamefont {Tillim}},\ }\bibfield
  {title} {\enquote {\bibinfo {title} {Lectures on generalized symmetries},}\
  }\href {\doibase https://doi.org/10.1016/j.physrep.2023.11.002} {\bibfield
  {journal} {\bibinfo  {journal} {Physics Reports}\ }\textbf {\bibinfo {volume}
  {1051}},\ \bibinfo {pages} {1} (\bibinfo {year} {2024}{\natexlab{a}})},\
  \Eprint {http://arxiv.org/abs/2307.07547} {arXiv:2307.07547 [hep-th]}
  \BibitemShut {NoStop}%
\bibitem [{\citenamefont {Gaiotto}\ \emph {et~al.}(2015)\citenamefont
  {Gaiotto}, \citenamefont {Kapustin}, \citenamefont {Seiberg},\ and\
  \citenamefont {Willett}}]{gaiotto2015generalized}%
  \BibitemOpen
  \bibfield  {author} {\bibinfo {author} {\bibfnamefont {D.}~\bibnamefont
  {Gaiotto}}, \bibinfo {author} {\bibfnamefont {A.}~\bibnamefont {Kapustin}},
  \bibinfo {author} {\bibfnamefont {N.}~\bibnamefont {Seiberg}}, \ and\
  \bibinfo {author} {\bibfnamefont {B.}~\bibnamefont {Willett}},\ }\bibfield
  {title} {\enquote {\bibinfo {title} {Generalized global symmetries},}\
  }\href@noop {} {\bibfield  {journal} {\bibinfo  {journal} {Journal of High
  Energy Physics}\ }\textbf {\bibinfo {volume} {2015}},\ \bibinfo {pages} {1}
  (\bibinfo {year} {2015})},\ \Eprint {http://arxiv.org/abs/1412.5148}
  {arXiv:1412.5148 [hep-th]} \BibitemShut {NoStop}%
\bibitem [{\citenamefont {Cao}\ \emph {et~al.}(2023)\citenamefont {Cao},
  \citenamefont {Li}, \citenamefont {Yamazaki},\ and\ \citenamefont
  {Zheng}}]{Cao2023subsystem}%
  \BibitemOpen
  \bibfield  {author} {\bibinfo {author} {\bibfnamefont {W.}~\bibnamefont
  {Cao}}, \bibinfo {author} {\bibfnamefont {L.}~\bibnamefont {Li}}, \bibinfo
  {author} {\bibfnamefont {M.}~\bibnamefont {Yamazaki}}, \ and\ \bibinfo
  {author} {\bibfnamefont {Y.}~\bibnamefont {Zheng}},\ }\bibfield  {title}
  {\enquote {\bibinfo {title} {{Subsystem non-invertible symmetry operators and
  defects}},}\ }\href {\doibase 10.21468/SciPostPhys.15.4.155} {\bibfield
  {journal} {\bibinfo  {journal} {SciPost Phys.}\ }\textbf {\bibinfo {volume}
  {15}},\ \bibinfo {pages} {155} (\bibinfo {year} {2023})},\ \Eprint
  {http://arxiv.org/abs/2304.09886} {arXiv:2304.09886 [cond-mat.str-el]}
  \BibitemShut {NoStop}%
\bibitem [{\citenamefont {Kapustin}\ and\ \citenamefont
  {Thorngren}(2017)}]{kapustin2017higher}%
  \BibitemOpen
  \bibfield  {author} {\bibinfo {author} {\bibfnamefont {A.}~\bibnamefont
  {Kapustin}}\ and\ \bibinfo {author} {\bibfnamefont {R.}~\bibnamefont
  {Thorngren}},\ }\bibfield  {title} {\enquote {\bibinfo {title} {Higher
  symmetry and gapped phases of gauge theories},}\ }\href
  {https://link-springer-com.libproxy1.nus.edu.sg/chapter/10.1007/978-3-319-59939-7_5}
  {\bibfield  {journal} {\bibinfo  {journal} {Algebra, Geometry, and Physics in
  the 21st Century: Kontsevich Festschrift}\ ,\ \bibinfo {pages} {177}}
  (\bibinfo {year} {2017})}\BibitemShut {NoStop}%
\bibitem [{\citenamefont {C{\'o}rdova}\ \emph {et~al.}(2019)\citenamefont
  {C{\'o}rdova}, \citenamefont {Dumitrescu},\ and\ \citenamefont
  {Intriligator}}]{Cordova2019highergroup}%
  \BibitemOpen
  \bibfield  {author} {\bibinfo {author} {\bibfnamefont {C.}~\bibnamefont
  {C{\'o}rdova}}, \bibinfo {author} {\bibfnamefont {T.~T.}\ \bibnamefont
  {Dumitrescu}}, \ and\ \bibinfo {author} {\bibfnamefont {K.}~\bibnamefont
  {Intriligator}},\ }\bibfield  {title} {\enquote {\bibinfo {title} {Exploring
  2-group global symmetries},}\ }\href {\doibase 10.1007/JHEP02(2019)184}
  {\bibfield  {journal} {\bibinfo  {journal} {Journal of High Energy Physics}\
  }\textbf {\bibinfo {volume} {2019}},\ \bibinfo {pages} {184} (\bibinfo {year}
  {2019})},\ \Eprint {http://arxiv.org/abs/1802.04790} {arXiv:1802.04790
  [hep-th]} \BibitemShut {NoStop}%
\bibitem [{\citenamefont {Buerschaper}\ \emph {et~al.}(2013)\citenamefont
  {Buerschaper}, \citenamefont {Mombelli}, \citenamefont {Christandl},\ and\
  \citenamefont {Aguado}}]{Buerschaper2013a}%
  \BibitemOpen
  \bibfield  {author} {\bibinfo {author} {\bibfnamefont {O.}~\bibnamefont
  {Buerschaper}}, \bibinfo {author} {\bibfnamefont {J.~M.}\ \bibnamefont
  {Mombelli}}, \bibinfo {author} {\bibfnamefont {M.}~\bibnamefont
  {Christandl}}, \ and\ \bibinfo {author} {\bibfnamefont {M.}~\bibnamefont
  {Aguado}},\ }\bibfield  {title} {\enquote {\bibinfo {title} {A hierarchy of
  topological tensor network states},}\ }\href {\doibase 10.1063/1.4773316}
  {\bibfield  {journal} {\bibinfo  {journal} {Journal of Mathematical Physics}\
  }\textbf {\bibinfo {volume} {54}},\ \bibinfo {pages} {012201} (\bibinfo
  {year} {2013})},\ \Eprint {http://arxiv.org/abs/1007.5283} {arXiv:1007.5283
  [cond-mat.str-el]} \BibitemShut {NoStop}%
\bibitem [{\citenamefont {Meusburger}(2017)}]{meusburger2017kitaev}%
  \BibitemOpen
  \bibfield  {author} {\bibinfo {author} {\bibfnamefont {C.}~\bibnamefont
  {Meusburger}},\ }\bibfield  {title} {\enquote {\bibinfo {title} {Kitaev
  lattice models as a {H}opf algebra gauge theory},}\ }\href
  {https://link.springer.com/article/10.1007%2Fs00220-017-2860-7} {\bibfield
  {journal} {\bibinfo  {journal} {Communications in Mathematical Physics}\
  }\textbf {\bibinfo {volume} {353}},\ \bibinfo {pages} {413} (\bibinfo {year}
  {2017})},\ \Eprint {http://arxiv.org/abs/1607.01144} {arXiv:1607.01144
  [math.QA]} \BibitemShut {NoStop}%
\bibitem [{\citenamefont {Yan}\ \emph {et~al.}(2022)\citenamefont {Yan},
  \citenamefont {Chen},\ and\ \citenamefont {Cui}}]{chen2021ribbon}%
  \BibitemOpen
  \bibfield  {author} {\bibinfo {author} {\bibfnamefont {B.}~\bibnamefont
  {Yan}}, \bibinfo {author} {\bibfnamefont {P.}~\bibnamefont {Chen}}, \ and\
  \bibinfo {author} {\bibfnamefont {S.}~\bibnamefont {Cui}},\ }\bibfield
  {title} {\enquote {\bibinfo {title} {Ribbon operators in the generalized
  {K}itaev quantum double model based on {H}opf algebras},}\ }\href
  {https://iopscience.iop.org/article/10.1088/1751-8121/ac552c/meta} {\bibfield
   {journal} {\bibinfo  {journal} {Journal of Physics A: Mathematical and
  Theoretical}\ } (\bibinfo {year} {2022})},\ \Eprint
  {http://arxiv.org/abs/2105.08202} {arXiv:2105.08202 [cond-mat.str-el]}
  \BibitemShut {NoStop}%
\bibitem [{\citenamefont {Jia}\ \emph {et~al.}(2023{\natexlab{a}})\citenamefont
  {Jia}, \citenamefont {Kaszlikowski},\ and\ \citenamefont
  {Tan}}]{jia2023boundary}%
  \BibitemOpen
  \bibfield  {author} {\bibinfo {author} {\bibfnamefont {Z.}~\bibnamefont
  {Jia}}, \bibinfo {author} {\bibfnamefont {D.}~\bibnamefont {Kaszlikowski}}, \
  and\ \bibinfo {author} {\bibfnamefont {S.}~\bibnamefont {Tan}},\ }\bibfield
  {title} {\enquote {\bibinfo {title} {Boundary and domain wall theories of 2d
  generalized quantum double model},}\ }\href
  {https://link.springer.com/article/10.1007/JHEP07(2023)160} {\bibfield
  {journal} {\bibinfo  {journal} {Journal of High Energy Physics}\ }\textbf
  {\bibinfo {volume} {2023}},\ \bibinfo {pages} {1} (\bibinfo {year}
  {2023}{\natexlab{a}})},\ \Eprint {http://arxiv.org/abs/2207.03970}
  {arXiv:2207.03970 [quant-ph]} \BibitemShut {NoStop}%
\bibitem [{\citenamefont {Jia}\ \emph {et~al.}(2023{\natexlab{b}})\citenamefont
  {Jia}, \citenamefont {Tan}, \citenamefont {Kaszlikowski},\ and\ \citenamefont
  {Chang}}]{Jia2023weak}%
  \BibitemOpen
  \bibfield  {author} {\bibinfo {author} {\bibfnamefont {Z.}~\bibnamefont
  {Jia}}, \bibinfo {author} {\bibfnamefont {S.}~\bibnamefont {Tan}}, \bibinfo
  {author} {\bibfnamefont {D.}~\bibnamefont {Kaszlikowski}}, \ and\ \bibinfo
  {author} {\bibfnamefont {L.}~\bibnamefont {Chang}},\ }\bibfield  {title}
  {\enquote {\bibinfo {title} {On weak {H}opf symmetry and weak {H}opf quantum
  double model},}\ }\href {\doibase 10.1007/s00220-023-04792-9} {\bibfield
  {journal} {\bibinfo  {journal} {Communications in Mathematical Physics}\
  }\textbf {\bibinfo {volume} {402}},\ \bibinfo {pages} {3045} (\bibinfo {year}
  {2023}{\natexlab{b}})},\ \Eprint {http://arxiv.org/abs/2302.08131}
  {arXiv:2302.08131 [hep-th]} \BibitemShut {NoStop}%
\bibitem [{\citenamefont {Jia}(2024{\natexlab{a}})}]{jia2024generalized}%
  \BibitemOpen
  \bibfield  {author} {\bibinfo {author} {\bibfnamefont {Z.}~\bibnamefont
  {Jia}},\ }\bibfield  {title} {\enquote {\bibinfo {title} {Generalized cluster
  states from hopf algebras: non-invertible symmetry and hopf tensor network
  representation},}\ }\href {\doibase 10.1007/JHEP09(2024)147} {\bibfield
  {journal} {\bibinfo  {journal} {Journal of High Energy Physics}\ }\textbf
  {\bibinfo {volume} {2024}},\ \bibinfo {pages} {147} (\bibinfo {year}
  {2024}{\natexlab{a}})},\ \Eprint {http://arxiv.org/abs/2405.09277}
  {arXiv:2405.09277 [quant-ph]} \BibitemShut {NoStop}%
\bibitem [{\citenamefont {Jia}\ \emph {et~al.}(2024)\citenamefont {Jia},
  \citenamefont {Tan},\ and\ \citenamefont {Kaszlikowski}}]{jia2024weakTube}%
  \BibitemOpen
  \bibfield  {author} {\bibinfo {author} {\bibfnamefont {Z.}~\bibnamefont
  {Jia}}, \bibinfo {author} {\bibfnamefont {S.}~\bibnamefont {Tan}}, \ and\
  \bibinfo {author} {\bibfnamefont {D.}~\bibnamefont {Kaszlikowski}},\
  }\bibfield  {title} {\enquote {\bibinfo {title} {Weak {H}opf symmetry and
  tube algebra of the generalized multifusion string-net model},}\ }\href
  {https://doi.org/10.1007/JHEP07(2024)207} {\bibfield  {journal} {\bibinfo
  {journal} {Journal of High Enegy Physics}\ }\textbf {\bibinfo {volume}
  {07}},\ \bibinfo {pages} {207} (\bibinfo {year} {2024})},\ \Eprint
  {http://arxiv.org/abs/2403.04446} {arXiv:2403.04446 [hep-th]} \BibitemShut
  {NoStop}%
\bibitem [{\citenamefont {Choi}\ \emph {et~al.}(2024)\citenamefont {Choi},
  \citenamefont {Lu},\ and\ \citenamefont {Sun}}]{Choi2024duality}%
  \BibitemOpen
  \bibfield  {author} {\bibinfo {author} {\bibfnamefont {Y.}~\bibnamefont
  {Choi}}, \bibinfo {author} {\bibfnamefont {D.-C.}\ \bibnamefont {Lu}}, \ and\
  \bibinfo {author} {\bibfnamefont {Z.}~\bibnamefont {Sun}},\ }\bibfield
  {title} {\enquote {\bibinfo {title} {Self-duality under gauging a
  non-invertible symmetry},}\ }\href {\doibase 10.1007/JHEP01(2024)142}
  {\bibfield  {journal} {\bibinfo  {journal} {Journal of High Energy Physics}\
  }\textbf {\bibinfo {volume} {2024}},\ \bibinfo {pages} {142} (\bibinfo {year}
  {2024})},\ \Eprint {http://arxiv.org/abs/2310.01474} {arXiv:2310.01474
  [hep-th]} \BibitemShut {NoStop}%
\bibitem [{\citenamefont
  {Jia}(2024{\natexlab{b}})}]{jia2024weakhopfnoninvertible}%
  \BibitemOpen
  \bibfield  {author} {\bibinfo {author} {\bibfnamefont {Z.}~\bibnamefont
  {Jia}},\ }\href {\doibase 10.48550/arXiv.2412.15336} {\enquote {\bibinfo
  {title} {Weak {H}opf non-invertible symmetry-protected topological spin
  liquid and lattice realization of (1+1){D} symmetry topological field
  theory},}\ } (\bibinfo {year} {2024}{\natexlab{b}}),\ \Eprint
  {http://arxiv.org/abs/2412.15336} {arXiv:2412.15336 [hep-th]} \BibitemShut
  {NoStop}%
\bibitem [{\citenamefont {Jia}(2024{\natexlab{c}})}]{jia2024quantumcluster}%
  \BibitemOpen
  \bibfield  {author} {\bibinfo {author} {\bibfnamefont {Z.}~\bibnamefont
  {Jia}},\ }\href {\doibase 10.48550/arXiv.2412.19657} {\enquote {\bibinfo
  {title} {Quantum cluster state model with haagerup fusion category
  symmetry},}\ } (\bibinfo {year} {2024}{\natexlab{c}}),\ \Eprint
  {http://arxiv.org/abs/2412.19657} {arXiv:2412.19657 [math.QA]} \BibitemShut
  {NoStop}%
\bibitem [{\citenamefont {Meng}\ \emph {et~al.}(2024)\citenamefont {Meng},
  \citenamefont {Yang}, \citenamefont {Lan},\ and\ \citenamefont
  {Gu}}]{meng2024noninvertiblespt}%
  \BibitemOpen
  \bibfield  {author} {\bibinfo {author} {\bibfnamefont {C.}~\bibnamefont
  {Meng}}, \bibinfo {author} {\bibfnamefont {X.}~\bibnamefont {Yang}}, \bibinfo
  {author} {\bibfnamefont {T.}~\bibnamefont {Lan}}, \ and\ \bibinfo {author}
  {\bibfnamefont {Z.}~\bibnamefont {Gu}},\ }\href
  {https://arxiv.org/abs/2412.20546} {\enquote {\bibinfo {title}
  {Non-invertible spts: an on-site realization of (1+1)d anomaly-free fusion
  category symmetry},}\ } (\bibinfo {year} {2024}),\ \Eprint
  {http://arxiv.org/abs/2412.20546} {arXiv:2412.20546 [cond-mat.str-el]}
  \BibitemShut {NoStop}%
\bibitem [{\citenamefont {Inamura}(2022)}]{inamura2022lattice}%
  \BibitemOpen
  \bibfield  {author} {\bibinfo {author} {\bibfnamefont {K.}~\bibnamefont
  {Inamura}},\ }\bibfield  {title} {\enquote {\bibinfo {title} {On lattice
  models of gapped phases with fusion category symmetries},}\ }\href
  {https://link.springer.com/article/10.1007/JHEP03(2022)036} {\bibfield
  {journal} {\bibinfo  {journal} {Journal of High Energy Physics}\ }\textbf
  {\bibinfo {volume} {2022}},\ \bibinfo {pages} {1} (\bibinfo {year} {2022})},\
  \Eprint {http://arxiv.org/abs/2110.12882} {arXiv:2110.12882
  [cond-mat.str-el]} \BibitemShut {NoStop}%
\bibitem [{\citenamefont {Inamura}(2023)}]{inamura2023fermionization}%
  \BibitemOpen
  \bibfield  {author} {\bibinfo {author} {\bibfnamefont {K.}~\bibnamefont
  {Inamura}},\ }\bibfield  {title} {\enquote {\bibinfo {title} {Fermionization
  of fusion category symmetries in 1+1 dimensions},}\ }\href
  {https://link.springer.com/article/10.1007/JHEP10(2023)101} {\bibfield
  {journal} {\bibinfo  {journal} {Journal of High Energy Physics}\ }\textbf
  {\bibinfo {volume} {2023}},\ \bibinfo {pages} {1} (\bibinfo {year} {2023})},\
  \Eprint {http://arxiv.org/abs/2206.13159} {arXiv:2206.13159
  [cond-mat.str-el]} \BibitemShut {NoStop}%
\bibitem [{\citenamefont {You}\ \emph {et~al.}(2018)\citenamefont {You},
  \citenamefont {Devakul}, \citenamefont {Burnell},\ and\ \citenamefont
  {Sondhi}}]{You2018SSPT}%
  \BibitemOpen
  \bibfield  {author} {\bibinfo {author} {\bibfnamefont {Y.}~\bibnamefont
  {You}}, \bibinfo {author} {\bibfnamefont {T.}~\bibnamefont {Devakul}},
  \bibinfo {author} {\bibfnamefont {F.~J.}\ \bibnamefont {Burnell}}, \ and\
  \bibinfo {author} {\bibfnamefont {S.~L.}\ \bibnamefont {Sondhi}},\ }\bibfield
   {title} {\enquote {\bibinfo {title} {Subsystem symmetry protected
  topological order},}\ }\href {\doibase 10.1103/PhysRevB.98.035112} {\bibfield
   {journal} {\bibinfo  {journal} {Phys. Rev. B}\ }\textbf {\bibinfo {volume}
  {98}},\ \bibinfo {pages} {035112} (\bibinfo {year} {2018})},\ \Eprint
  {http://arxiv.org/abs/1803.02369} {arXiv:1803.02369 [cond-mat.str-el]}
  \BibitemShut {NoStop}%
\bibitem [{\citenamefont {Devakul}\ \emph {et~al.}(2018)\citenamefont
  {Devakul}, \citenamefont {Williamson},\ and\ \citenamefont
  {You}}]{Devakul2018classifcation}%
  \BibitemOpen
  \bibfield  {author} {\bibinfo {author} {\bibfnamefont {T.}~\bibnamefont
  {Devakul}}, \bibinfo {author} {\bibfnamefont {D.~J.}\ \bibnamefont
  {Williamson}}, \ and\ \bibinfo {author} {\bibfnamefont {Y.}~\bibnamefont
  {You}},\ }\bibfield  {title} {\enquote {\bibinfo {title} {Classification of
  subsystem symmetry-protected topological phases},}\ }\href {\doibase
  10.1103/PhysRevB.98.235121} {\bibfield  {journal} {\bibinfo  {journal} {Phys.
  Rev. B}\ }\textbf {\bibinfo {volume} {98}},\ \bibinfo {pages} {235121}
  (\bibinfo {year} {2018})},\ \Eprint {http://arxiv.org/abs/1808.05300}
  {arXiv:1808.05300 [cond-mat.str-el]} \BibitemShut {NoStop}%
\bibitem [{\citenamefont {Shirley}\ \emph {et~al.}(2019)\citenamefont
  {Shirley}, \citenamefont {Slagle},\ and\ \citenamefont
  {Chen}}]{Shirley2019foliated}%
  \BibitemOpen
  \bibfield  {author} {\bibinfo {author} {\bibfnamefont {W.}~\bibnamefont
  {Shirley}}, \bibinfo {author} {\bibfnamefont {K.}~\bibnamefont {Slagle}}, \
  and\ \bibinfo {author} {\bibfnamefont {X.}~\bibnamefont {Chen}},\ }\bibfield
  {title} {\enquote {\bibinfo {title} {{Foliated fracton order from gauging
  subsystem symmetries}},}\ }\href {\doibase 10.21468/SciPostPhys.6.4.041}
  {\bibfield  {journal} {\bibinfo  {journal} {SciPost Phys.}\ }\textbf
  {\bibinfo {volume} {6}},\ \bibinfo {pages} {041} (\bibinfo {year} {2019})},\
  \Eprint {http://arxiv.org/abs/1806.08679} {arXiv:1806.08679
  [cond-mat.str-el]} \BibitemShut {NoStop}%
\bibitem [{\citenamefont {Jia}\ and\ \citenamefont
  {Jia}(2025)}]{jia2025subsystemsymmetry}%
  \BibitemOpen
  \bibfield  {author} {\bibinfo {author} {\bibfnamefont {Q.}~\bibnamefont
  {Jia}}\ and\ \bibinfo {author} {\bibfnamefont {Z.}~\bibnamefont {Jia}},\
  }\href {https://arxiv.org/abs/2505.22261} {\enquote {\bibinfo {title}
  {Subsystem symmetry-protected topological phases from subsystem symtft of
  2-foliated exotic tensor gauge theory},}\ } (\bibinfo {year} {2025}),\
  \Eprint {http://arxiv.org/abs/2505.22261} {arXiv:2505.22261
  [cond-mat.str-el]} \BibitemShut {NoStop}%
\bibitem [{\citenamefont {Kong}\ \emph {et~al.}(2020)\citenamefont {Kong},
  \citenamefont {Lan}, \citenamefont {Wen}, \citenamefont {Zhang},\ and\
  \citenamefont {Zheng}}]{Kong2020algebraic}%
  \BibitemOpen
  \bibfield  {author} {\bibinfo {author} {\bibfnamefont {L.}~\bibnamefont
  {Kong}}, \bibinfo {author} {\bibfnamefont {T.}~\bibnamefont {Lan}}, \bibinfo
  {author} {\bibfnamefont {X.-G.}\ \bibnamefont {Wen}}, \bibinfo {author}
  {\bibfnamefont {Z.-H.}\ \bibnamefont {Zhang}}, \ and\ \bibinfo {author}
  {\bibfnamefont {H.}~\bibnamefont {Zheng}},\ }\bibfield  {title} {\enquote
  {\bibinfo {title} {Algebraic higher symmetry and categorical symmetry: A
  holographic and entanglement view of symmetry},}\ }\href {\doibase
  10.1103/PhysRevResearch.2.043086} {\bibfield  {journal} {\bibinfo  {journal}
  {Phys. Rev. Res.}\ }\textbf {\bibinfo {volume} {2}},\ \bibinfo {pages}
  {043086} (\bibinfo {year} {2020})},\ \Eprint
  {http://arxiv.org/abs/2005.14178} {arXiv:2005.14178 [cond-mat.str-el]}
  \BibitemShut {NoStop}%
\bibitem [{\citenamefont {Huang}\ and\ \citenamefont
  {Cheng}(2023)}]{huang2023topologicalholo}%
  \BibitemOpen
  \bibfield  {author} {\bibinfo {author} {\bibfnamefont {S.-J.}\ \bibnamefont
  {Huang}}\ and\ \bibinfo {author} {\bibfnamefont {M.}~\bibnamefont {Cheng}},\
  }\href {https://arxiv.org/abs/2310.16878} {\enquote {\bibinfo {title}
  {Topological holography, quantum criticality, and boundary states},}\ }
  (\bibinfo {year} {2023}),\ \Eprint {http://arxiv.org/abs/2310.16878}
  {arXiv:2310.16878 [cond-mat.str-el]} \BibitemShut {NoStop}%
\bibitem [{\citenamefont {Bhardwaj}\ \emph
  {et~al.}(2024{\natexlab{b}})\citenamefont {Bhardwaj}, \citenamefont
  {Bottini}, \citenamefont {Schafer-Nameki},\ and\ \citenamefont
  {Tiwari}}]{bhardwaj2024lattice}%
  \BibitemOpen
  \bibfield  {author} {\bibinfo {author} {\bibfnamefont {L.}~\bibnamefont
  {Bhardwaj}}, \bibinfo {author} {\bibfnamefont {L.~E.}\ \bibnamefont
  {Bottini}}, \bibinfo {author} {\bibfnamefont {S.}~\bibnamefont
  {Schafer-Nameki}}, \ and\ \bibinfo {author} {\bibfnamefont {A.}~\bibnamefont
  {Tiwari}},\ }\href@noop {} {\enquote {\bibinfo {title} {Lattice models for
  phases and transitions with non-invertible symmetries},}\ } (\bibinfo {year}
  {2024}{\natexlab{b}}),\ \Eprint {http://arxiv.org/abs/2405.05964}
  {arXiv:2405.05964 [cond-mat.str-el]} \BibitemShut {NoStop}%
\bibitem [{\citenamefont {Freed}\ \emph {et~al.}(2024)\citenamefont {Freed},
  \citenamefont {Moore},\ and\ \citenamefont {Teleman}}]{freed2024topSymTFT}%
  \BibitemOpen
  \bibfield  {author} {\bibinfo {author} {\bibfnamefont {D.~S.}\ \bibnamefont
  {Freed}}, \bibinfo {author} {\bibfnamefont {G.~W.}\ \bibnamefont {Moore}}, \
  and\ \bibinfo {author} {\bibfnamefont {C.}~\bibnamefont {Teleman}},\ }\href
  {https://arxiv.org/abs/2209.07471} {\enquote {\bibinfo {title} {Topological
  symmetry in quantum field theory},}\ } (\bibinfo {year} {2024}),\ \Eprint
  {http://arxiv.org/abs/2209.07471} {arXiv:2209.07471 [hep-th]} \BibitemShut
  {NoStop}%
\bibitem [{\citenamefont {Gaiotto}\ and\ \citenamefont
  {Kulp}(2021)}]{gaiotto2021orbifold}%
  \BibitemOpen
  \bibfield  {author} {\bibinfo {author} {\bibfnamefont {D.}~\bibnamefont
  {Gaiotto}}\ and\ \bibinfo {author} {\bibfnamefont {J.}~\bibnamefont {Kulp}},\
  }\bibfield  {title} {\enquote {\bibinfo {title} {Orbifold groupoids},}\
  }\href {https://link.springer.com/article/10.1007/JHEP02(2021)132} {\bibfield
   {journal} {\bibinfo  {journal} {Journal of High Energy Physics}\ }\textbf
  {\bibinfo {volume} {2021}},\ \bibinfo {pages} {1} (\bibinfo {year} {2021})},\
  \Eprint {http://arxiv.org/abs/2008.05960} {arXiv:2008.05960 [hep-th]}
  \BibitemShut {NoStop}%
\bibitem [{\citenamefont {Bhardwaj}\ and\ \citenamefont
  {Schafer-Nameki}(2023)}]{bhardwaj2023generalizedcharge}%
  \BibitemOpen
  \bibfield  {author} {\bibinfo {author} {\bibfnamefont {L.}~\bibnamefont
  {Bhardwaj}}\ and\ \bibinfo {author} {\bibfnamefont {S.}~\bibnamefont
  {Schafer-Nameki}},\ }\href {https://arxiv.org/abs/2305.17159} {\enquote
  {\bibinfo {title} {Generalized charges, part ii: Non-invertible symmetries
  and the symmetry {TFT}},}\ } (\bibinfo {year} {2023}),\ \Eprint
  {http://arxiv.org/abs/2305.17159} {arXiv:2305.17159 [hep-th]} \BibitemShut
  {NoStop}%
\bibitem [{\citenamefont {Apruzzi}\ \emph {et~al.}(2023)\citenamefont
  {Apruzzi}, \citenamefont {Bonetti}, \citenamefont {Garc{\'\i}a~Etxebarria},
  \citenamefont {Hosseini},\ and\ \citenamefont
  {Sch{\"a}fer-Nameki}}]{apruzzi2023symmetry}%
  \BibitemOpen
  \bibfield  {author} {\bibinfo {author} {\bibfnamefont {F.}~\bibnamefont
  {Apruzzi}}, \bibinfo {author} {\bibfnamefont {F.}~\bibnamefont {Bonetti}},
  \bibinfo {author} {\bibfnamefont {I.}~\bibnamefont {Garc{\'\i}a~Etxebarria}},
  \bibinfo {author} {\bibfnamefont {S.~S.}\ \bibnamefont {Hosseini}}, \ and\
  \bibinfo {author} {\bibfnamefont {S.}~\bibnamefont {Sch{\"a}fer-Nameki}},\
  }\bibfield  {title} {\enquote {\bibinfo {title} {Symmetry tfts from string
  theory},}\ }\href
  {https://link.springer.com/article/10.1007/s00220-023-04737-2} {\bibfield
  {journal} {\bibinfo  {journal} {Communications in mathematical physics}\
  }\textbf {\bibinfo {volume} {402}},\ \bibinfo {pages} {895} (\bibinfo {year}
  {2023})},\ \Eprint {http://arxiv.org/abs/2112.02092} {arXiv:2112.02092
  [hep-th]} \BibitemShut {NoStop}%
\bibitem [{\citenamefont {Bhardwaj}\ \emph
  {et~al.}(2024{\natexlab{c}})\citenamefont {Bhardwaj}, \citenamefont
  {Bottini}, \citenamefont {Pajer},\ and\ \citenamefont
  {Schafer-Nameki}}]{bhardwaj2024gappedphases}%
  \BibitemOpen
  \bibfield  {author} {\bibinfo {author} {\bibfnamefont {L.}~\bibnamefont
  {Bhardwaj}}, \bibinfo {author} {\bibfnamefont {L.~E.}\ \bibnamefont
  {Bottini}}, \bibinfo {author} {\bibfnamefont {D.}~\bibnamefont {Pajer}}, \
  and\ \bibinfo {author} {\bibfnamefont {S.}~\bibnamefont {Schafer-Nameki}},\
  }\href {https://arxiv.org/abs/2310.03784} {\enquote {\bibinfo {title} {Gapped
  phases with non-invertible symmetries: (1+1)d},}\ } (\bibinfo {year}
  {2024}{\natexlab{c}}),\ \Eprint {http://arxiv.org/abs/2310.03784}
  {arXiv:2310.03784 [hep-th]} \BibitemShut {NoStop}%
\bibitem [{\citenamefont {Zhang}\ and\ \citenamefont
  {C\'ordova}(2024)}]{Zhang2024anomaly}%
  \BibitemOpen
  \bibfield  {author} {\bibinfo {author} {\bibfnamefont {C.}~\bibnamefont
  {Zhang}}\ and\ \bibinfo {author} {\bibfnamefont {C.}~\bibnamefont
  {C\'ordova}},\ }\bibfield  {title} {\enquote {\bibinfo {title} {Anomalies of
  $(1+1)$-dimensional categorical symmetries},}\ }\href {\doibase
  10.1103/PhysRevB.110.035155} {\bibfield  {journal} {\bibinfo  {journal}
  {Phys. Rev. B}\ }\textbf {\bibinfo {volume} {110}},\ \bibinfo {pages}
  {035155} (\bibinfo {year} {2024})}\BibitemShut {NoStop}%
\bibitem [{\citenamefont {Ji}\ and\ \citenamefont
  {Wen}(2020)}]{Ji2020categoricalsym}%
  \BibitemOpen
  \bibfield  {author} {\bibinfo {author} {\bibfnamefont {W.}~\bibnamefont
  {Ji}}\ and\ \bibinfo {author} {\bibfnamefont {X.-G.}\ \bibnamefont {Wen}},\
  }\bibfield  {title} {\enquote {\bibinfo {title} {Categorical symmetry and
  noninvertible anomaly in symmetry-breaking and topological phase
  transitions},}\ }\href {\doibase 10.1103/PhysRevResearch.2.033417} {\bibfield
   {journal} {\bibinfo  {journal} {Phys. Rev. Res.}\ }\textbf {\bibinfo
  {volume} {2}},\ \bibinfo {pages} {033417} (\bibinfo {year}
  {2020})}\BibitemShut {NoStop}%
\bibitem [{\citenamefont {Bhardwaj}\ \emph
  {et~al.}(2024{\natexlab{d}})\citenamefont {Bhardwaj}, \citenamefont
  {Bottini}, \citenamefont {Pajer},\ and\ \citenamefont
  {Schafer-Nameki}}]{bhardwaj2024clubsandwich}%
  \BibitemOpen
  \bibfield  {author} {\bibinfo {author} {\bibfnamefont {L.}~\bibnamefont
  {Bhardwaj}}, \bibinfo {author} {\bibfnamefont {L.~E.}\ \bibnamefont
  {Bottini}}, \bibinfo {author} {\bibfnamefont {D.}~\bibnamefont {Pajer}}, \
  and\ \bibinfo {author} {\bibfnamefont {S.}~\bibnamefont {Schafer-Nameki}},\
  }\href {https://arxiv.org/abs/2312.17322} {\enquote {\bibinfo {title} {The
  club sandwich: Gapless phases and phase transitions with non-invertible
  symmetries},}\ } (\bibinfo {year} {2024}{\natexlab{d}}),\ \Eprint
  {http://arxiv.org/abs/2312.17322} {arXiv:2312.17322 [hep-th]} \BibitemShut
  {NoStop}%
\bibitem [{\citenamefont {Bhardwaj}\ \emph
  {et~al.}(2024{\natexlab{e}})\citenamefont {Bhardwaj}, \citenamefont {Pajer},
  \citenamefont {Schafer-Nameki},\ and\ \citenamefont
  {Warman}}]{bhardwaj2024hassediagramsgaplessspt}%
  \BibitemOpen
  \bibfield  {author} {\bibinfo {author} {\bibfnamefont {L.}~\bibnamefont
  {Bhardwaj}}, \bibinfo {author} {\bibfnamefont {D.}~\bibnamefont {Pajer}},
  \bibinfo {author} {\bibfnamefont {S.}~\bibnamefont {Schafer-Nameki}}, \ and\
  \bibinfo {author} {\bibfnamefont {A.}~\bibnamefont {Warman}},\ }\href
  {https://arxiv.org/abs/2403.00905} {\enquote {\bibinfo {title} {Hasse
  diagrams for gapless spt and ssb phases with non-invertible symmetries},}\ }
  (\bibinfo {year} {2024}{\natexlab{e}}),\ \Eprint
  {http://arxiv.org/abs/2403.00905} {arXiv:2403.00905 [cond-mat.str-el]}
  \BibitemShut {NoStop}%
\bibitem [{\citenamefont {Thorngren}\ and\ \citenamefont
  {Wang}(2024)}]{thorngren2019fusion}%
  \BibitemOpen
  \bibfield  {author} {\bibinfo {author} {\bibfnamefont {R.}~\bibnamefont
  {Thorngren}}\ and\ \bibinfo {author} {\bibfnamefont {Y.}~\bibnamefont
  {Wang}},\ }\bibfield  {title} {\enquote {\bibinfo {title} {Fusion category
  symmetry. part i. anomaly in-flow and gapped phases},}\ }\href
  {https://link.springer.com/article/10.1007/JHEP04(2024)132} {\bibfield
  {journal} {\bibinfo  {journal} {Journal of High Energy Physics}\ }\textbf
  {\bibinfo {volume} {2024}},\ \bibinfo {pages} {1} (\bibinfo {year} {2024})},\
  \Eprint {http://arxiv.org/abs/1912.02817} {arXiv:1912.02817 [hep-th]}
  \BibitemShut {NoStop}%
\bibitem [{\citenamefont {Aasen}\ \emph {et~al.}(2016)\citenamefont {Aasen},
  \citenamefont {Mong},\ and\ \citenamefont {Fendley}}]{aasen2016topological}%
  \BibitemOpen
  \bibfield  {author} {\bibinfo {author} {\bibfnamefont {D.}~\bibnamefont
  {Aasen}}, \bibinfo {author} {\bibfnamefont {R.~S.}\ \bibnamefont {Mong}}, \
  and\ \bibinfo {author} {\bibfnamefont {P.}~\bibnamefont {Fendley}},\
  }\bibfield  {title} {\enquote {\bibinfo {title} {Topological defects on the
  lattice: I. {T}he {I}sing model},}\ }\href
  {https://iopscience.iop.org/article/10.1088/1751-8113/49/35/354001}
  {\bibfield  {journal} {\bibinfo  {journal} {Journal of Physics A:
  Mathematical and Theoretical}\ }\textbf {\bibinfo {volume} {49}},\ \bibinfo
  {pages} {354001} (\bibinfo {year} {2016})},\ \Eprint
  {http://arxiv.org/abs/1601.07185} {arXiv:1601.07185 [cond-mat.stat-mech]}
  \BibitemShut {NoStop}%
\bibitem [{\citenamefont {Aasen}\ \emph {et~al.}(2020)\citenamefont {Aasen},
  \citenamefont {Fendley},\ and\ \citenamefont {Mong}}]{aasen2020topological}%
  \BibitemOpen
  \bibfield  {author} {\bibinfo {author} {\bibfnamefont {D.}~\bibnamefont
  {Aasen}}, \bibinfo {author} {\bibfnamefont {P.}~\bibnamefont {Fendley}}, \
  and\ \bibinfo {author} {\bibfnamefont {R.~S.}\ \bibnamefont {Mong}},\
  }\bibfield  {title} {\enquote {\bibinfo {title} {Topological defects on the
  lattice: dualities and degeneracies},}\ }\href
  {https://arxiv.org/abs/2008.08598} {\bibfield  {journal} {\bibinfo  {journal}
  {arXiv preprint arXiv:2008.08598}\ } (\bibinfo {year} {2020})}\BibitemShut
  {NoStop}%
\bibitem [{\citenamefont {Inamura}\ and\ \citenamefont
  {Ohmori}(2024)}]{Inamura2024fusionSuface}%
  \BibitemOpen
  \bibfield  {author} {\bibinfo {author} {\bibfnamefont {K.}~\bibnamefont
  {Inamura}}\ and\ \bibinfo {author} {\bibfnamefont {K.}~\bibnamefont
  {Ohmori}},\ }\bibfield  {title} {\enquote {\bibinfo {title} {{Fusion surface
  models: 2+1d lattice models from fusion 2-categories}},}\ }\href {\doibase
  10.21468/SciPostPhys.16.6.143} {\bibfield  {journal} {\bibinfo  {journal}
  {SciPost Phys.}\ }\textbf {\bibinfo {volume} {16}},\ \bibinfo {pages} {143}
  (\bibinfo {year} {2024})},\ \Eprint {http://arxiv.org/abs/2305.05774}
  {arXiv:2305.05774 [cond-mat.str-el]} \BibitemShut {NoStop}%
\bibitem [{\citenamefont {Eck}\ and\ \citenamefont
  {Fendley}(2024)}]{eck2024generalizationskitaevshoneycombmodel}%
  \BibitemOpen
  \bibfield  {author} {\bibinfo {author} {\bibfnamefont {L.}~\bibnamefont
  {Eck}}\ and\ \bibinfo {author} {\bibfnamefont {P.}~\bibnamefont {Fendley}},\
  }\href {https://arxiv.org/abs/2408.04006} {\enquote {\bibinfo {title}
  {Generalizations of kitaev's honeycomb model from braided fusion
  categories},}\ } (\bibinfo {year} {2024}),\ \Eprint
  {http://arxiv.org/abs/2408.04006} {arXiv:2408.04006 [cond-mat.str-el]}
  \BibitemShut {NoStop}%
\bibitem [{\citenamefont {Fechisin}\ \emph {et~al.}(2025)\citenamefont
  {Fechisin}, \citenamefont {Tantivasadakarn},\ and\ \citenamefont
  {Albert}}]{fechisin2023noninvertible}%
  \BibitemOpen
  \bibfield  {author} {\bibinfo {author} {\bibfnamefont {C.}~\bibnamefont
  {Fechisin}}, \bibinfo {author} {\bibfnamefont {N.}~\bibnamefont
  {Tantivasadakarn}}, \ and\ \bibinfo {author} {\bibfnamefont {V.~V.}\
  \bibnamefont {Albert}},\ }\bibfield  {title} {\enquote {\bibinfo {title}
  {Noninvertible symmetry-protected topological order in a group-based cluster
  state},}\ }\href {\doibase 10.1103/PhysRevX.15.011058} {\bibfield  {journal}
  {\bibinfo  {journal} {Phys. Rev. X}\ }\textbf {\bibinfo {volume} {15}},\
  \bibinfo {pages} {011058} (\bibinfo {year} {2025})},\ \Eprint
  {http://arxiv.org/abs/2312.09272} {arXiv:2312.09272 [cond-mat.str-el]}
  \BibitemShut {NoStop}%
\bibitem [{\citenamefont {Feiguin}\ \emph {et~al.}(2007)\citenamefont
  {Feiguin}, \citenamefont {Trebst}, \citenamefont {Ludwig}, \citenamefont
  {Troyer}, \citenamefont {Kitaev}, \citenamefont {Wang},\ and\ \citenamefont
  {Freedman}}]{Feiguin2007interacting}%
  \BibitemOpen
  \bibfield  {author} {\bibinfo {author} {\bibfnamefont {A.}~\bibnamefont
  {Feiguin}}, \bibinfo {author} {\bibfnamefont {S.}~\bibnamefont {Trebst}},
  \bibinfo {author} {\bibfnamefont {A.~W.~W.}\ \bibnamefont {Ludwig}}, \bibinfo
  {author} {\bibfnamefont {M.}~\bibnamefont {Troyer}}, \bibinfo {author}
  {\bibfnamefont {A.}~\bibnamefont {Kitaev}}, \bibinfo {author} {\bibfnamefont
  {Z.}~\bibnamefont {Wang}}, \ and\ \bibinfo {author} {\bibfnamefont {M.~H.}\
  \bibnamefont {Freedman}},\ }\bibfield  {title} {\enquote {\bibinfo {title}
  {Interacting anyons in topological quantum liquids: The golden chain},}\
  }\href {\doibase 10.1103/PhysRevLett.98.160409} {\bibfield  {journal}
  {\bibinfo  {journal} {Phys. Rev. Lett.}\ }\textbf {\bibinfo {volume} {98}},\
  \bibinfo {pages} {160409} (\bibinfo {year} {2007})},\ \Eprint
  {http://arxiv.org/abs/cond-mat/0612341} {arXiv:cond-mat/0612341
  [cond-mat.str-el]} \BibitemShut {NoStop}%
\bibitem [{\citenamefont {Trebst}\ \emph {et~al.}(2008)\citenamefont {Trebst},
  \citenamefont {Troyer}, \citenamefont {Wang},\ and\ \citenamefont
  {Ludwig}}]{trebst2008short}%
  \BibitemOpen
  \bibfield  {author} {\bibinfo {author} {\bibfnamefont {S.}~\bibnamefont
  {Trebst}}, \bibinfo {author} {\bibfnamefont {M.}~\bibnamefont {Troyer}},
  \bibinfo {author} {\bibfnamefont {Z.}~\bibnamefont {Wang}}, \ and\ \bibinfo
  {author} {\bibfnamefont {A.~W.}\ \bibnamefont {Ludwig}},\ }\bibfield  {title}
  {\enquote {\bibinfo {title} {A short introduction to fibonacci anyon
  models},}\ }\href@noop {} {\bibfield  {journal} {\bibinfo  {journal}
  {Progress of Theoretical Physics Supplement}\ }\textbf {\bibinfo {volume}
  {176}},\ \bibinfo {pages} {384} (\bibinfo {year} {2008})}\BibitemShut
  {NoStop}%
\bibitem [{\citenamefont {Buican}\ and\ \citenamefont
  {Gromov}(2017)}]{buican2017anyonic}%
  \BibitemOpen
  \bibfield  {author} {\bibinfo {author} {\bibfnamefont {M.}~\bibnamefont
  {Buican}}\ and\ \bibinfo {author} {\bibfnamefont {A.}~\bibnamefont
  {Gromov}},\ }\bibfield  {title} {\enquote {\bibinfo {title} {Anyonic chains,
  topological defects, and conformal field theory},}\ }\href
  {https://link.springer.com/article/10.1007/s00220-017-2995-6} {\bibfield
  {journal} {\bibinfo  {journal} {Communications in Mathematical Physics}\
  }\textbf {\bibinfo {volume} {356}},\ \bibinfo {pages} {1017} (\bibinfo {year}
  {2017})},\ \Eprint {http://arxiv.org/abs/1701.02800} {arXiv:1701.02800
  [hep-th]} \BibitemShut {NoStop}%
\bibitem [{\citenamefont {Lootens}\ \emph {et~al.}(2023)\citenamefont
  {Lootens}, \citenamefont {Delcamp}, \citenamefont {Ortiz},\ and\
  \citenamefont {Verstraete}}]{Lootens2023dualityHamiltonian}%
  \BibitemOpen
  \bibfield  {author} {\bibinfo {author} {\bibfnamefont {L.}~\bibnamefont
  {Lootens}}, \bibinfo {author} {\bibfnamefont {C.}~\bibnamefont {Delcamp}},
  \bibinfo {author} {\bibfnamefont {G.}~\bibnamefont {Ortiz}}, \ and\ \bibinfo
  {author} {\bibfnamefont {F.}~\bibnamefont {Verstraete}},\ }\bibfield  {title}
  {\enquote {\bibinfo {title} {Dualities in one-dimensional quantum lattice
  models: Symmetric hamiltonians and matrix product operator intertwiners},}\
  }\href {\doibase 10.1103/PRXQuantum.4.020357} {\bibfield  {journal} {\bibinfo
   {journal} {PRX Quantum}\ }\textbf {\bibinfo {volume} {4}},\ \bibinfo {pages}
  {020357} (\bibinfo {year} {2023})},\ \Eprint
  {http://arxiv.org/abs/2112.09091} {arXiv:2112.09091 [quant-ph]} \BibitemShut
  {NoStop}%
\bibitem [{\citenamefont {Lootens}\ \emph {et~al.}(2024)\citenamefont
  {Lootens}, \citenamefont {Delcamp},\ and\ \citenamefont
  {Verstraete}}]{Lootens2024duality}%
  \BibitemOpen
  \bibfield  {author} {\bibinfo {author} {\bibfnamefont {L.}~\bibnamefont
  {Lootens}}, \bibinfo {author} {\bibfnamefont {C.}~\bibnamefont {Delcamp}}, \
  and\ \bibinfo {author} {\bibfnamefont {F.}~\bibnamefont {Verstraete}},\
  }\bibfield  {title} {\enquote {\bibinfo {title} {Dualities in one-dimensional
  quantum lattice models: Topological sectors},}\ }\href {\doibase
  10.1103/PRXQuantum.5.010338} {\bibfield  {journal} {\bibinfo  {journal} {PRX
  Quantum}\ }\textbf {\bibinfo {volume} {5}},\ \bibinfo {pages} {010338}
  (\bibinfo {year} {2024})},\ \Eprint {http://arxiv.org/abs/2211.03777}
  {arXiv:2211.03777 [quant-ph]} \BibitemShut {NoStop}%
\bibitem [{\citenamefont {Seiberg}\ \emph {et~al.}(2024)\citenamefont
  {Seiberg}, \citenamefont {Seifnashri},\ and\ \citenamefont
  {Shao}}]{Seiberg2024noninvertible}%
  \BibitemOpen
  \bibfield  {author} {\bibinfo {author} {\bibfnamefont {N.}~\bibnamefont
  {Seiberg}}, \bibinfo {author} {\bibfnamefont {S.}~\bibnamefont {Seifnashri}},
  \ and\ \bibinfo {author} {\bibfnamefont {S.-H.}\ \bibnamefont {Shao}},\
  }\bibfield  {title} {\enquote {\bibinfo {title} {{Non-invertible symmetries
  and LSM-type constraints on a tensor product Hilbert space}},}\ }\href
  {\doibase 10.21468/SciPostPhys.16.6.154} {\bibfield  {journal} {\bibinfo
  {journal} {SciPost Phys.}\ }\textbf {\bibinfo {volume} {16}},\ \bibinfo
  {pages} {154} (\bibinfo {year} {2024})},\ \Eprint
  {http://arxiv.org/abs/2401.12281} {arXiv:2401.12281 [cond-mat.str-el]}
  \BibitemShut {NoStop}%
\bibitem [{\citenamefont {Seifnashri}\ and\ \citenamefont
  {Shao}(2024)}]{Seifnashri2024cluster}%
  \BibitemOpen
  \bibfield  {author} {\bibinfo {author} {\bibfnamefont {S.}~\bibnamefont
  {Seifnashri}}\ and\ \bibinfo {author} {\bibfnamefont {S.-H.}\ \bibnamefont
  {Shao}},\ }\bibfield  {title} {\enquote {\bibinfo {title} {Cluster state as a
  noninvertible symmetry-protected topological phase},}\ }\href {\doibase
  10.1103/PhysRevLett.133.116601} {\bibfield  {journal} {\bibinfo  {journal}
  {Phys. Rev. Lett.}\ }\textbf {\bibinfo {volume} {133}},\ \bibinfo {pages}
  {116601} (\bibinfo {year} {2024})},\ \Eprint
  {http://arxiv.org/abs/2404.01369} {arXiv:2404.01369 [cond-mat.str-el]}
  \BibitemShut {NoStop}%
\bibitem [{\citenamefont {Inamura}\ and\ \citenamefont
  {Ohyama}(2024)}]{inamura202411dsptphasesfusion}%
  \BibitemOpen
  \bibfield  {author} {\bibinfo {author} {\bibfnamefont {K.}~\bibnamefont
  {Inamura}}\ and\ \bibinfo {author} {\bibfnamefont {S.}~\bibnamefont
  {Ohyama}},\ }\href {https://arxiv.org/abs/2408.15960} {\enquote {\bibinfo
  {title} {1+1d spt phases with fusion category symmetry: interface modes and
  non-abelian thouless pump},}\ } (\bibinfo {year} {2024}),\ \Eprint
  {http://arxiv.org/abs/2408.15960} {arXiv:2408.15960 [cond-mat.str-el]}
  \BibitemShut {NoStop}%
\bibitem [{\citenamefont {Cordova}\ \emph {et~al.}(2024)\citenamefont
  {Cordova}, \citenamefont {Costa},\ and\ \citenamefont
  {Hsin}}]{cordova2024noninvertiblesymmetriesfinitegroup}%
  \BibitemOpen
  \bibfield  {author} {\bibinfo {author} {\bibfnamefont {C.}~\bibnamefont
  {Cordova}}, \bibinfo {author} {\bibfnamefont {D.~B.}\ \bibnamefont {Costa}},
  \ and\ \bibinfo {author} {\bibfnamefont {P.-S.}\ \bibnamefont {Hsin}},\
  }\href {https://arxiv.org/abs/2407.07964} {\enquote {\bibinfo {title}
  {Non-invertible symmetries in finite group gauge theory},}\ } (\bibinfo
  {year} {2024}),\ \Eprint {http://arxiv.org/abs/2407.07964} {arXiv:2407.07964
  [cond-mat.str-el]} \BibitemShut {NoStop}%
\bibitem [{\citenamefont {Jia}\ and\ \citenamefont
  {Tan}(2025)}]{jia2025weakhopftubealgebra}%
  \BibitemOpen
  \bibfield  {author} {\bibinfo {author} {\bibfnamefont {Z.}~\bibnamefont
  {Jia}}\ and\ \bibinfo {author} {\bibfnamefont {S.}~\bibnamefont {Tan}},\
  }\href {https://arxiv.org/abs/2507.01515} {\enquote {\bibinfo {title} {Weak
  hopf tube algebra for domain walls between 2d gapped phases of turaev-viro
  tqfts},}\ } (\bibinfo {year} {2025}),\ \Eprint
  {http://arxiv.org/abs/2507.01515} {arXiv:2507.01515 [hep-th]} \BibitemShut
  {NoStop}%
\bibitem [{\citenamefont {Ketov}(1995)}]{ketov1995conformal}%
  \BibitemOpen
  \bibfield  {author} {\bibinfo {author} {\bibfnamefont {S.~V.}\ \bibnamefont
  {Ketov}},\ }\href@noop {} {\emph {\bibinfo {title} {Conformal field
  theory}}}\ (\bibinfo  {publisher} {World Scientific},\ \bibinfo {year}
  {1995})\BibitemShut {NoStop}%
\bibitem [{\citenamefont {Bakalov}\ and\ \citenamefont
  {Kirillov}(2001)}]{bakalov2001lectures}%
  \BibitemOpen
  \bibfield  {author} {\bibinfo {author} {\bibfnamefont {B.}~\bibnamefont
  {Bakalov}}\ and\ \bibinfo {author} {\bibfnamefont {A.~A.}\ \bibnamefont
  {Kirillov}},\ }\href {http://bookstore.ams.org/ulect-21} {\emph {\bibinfo
  {title} {Lectures on tensor categories and modular functors}}},\
  Vol.~\bibinfo {volume} {21}\ (\bibinfo  {publisher} {American Mathematical
  Soc.},\ \bibinfo {year} {2001})\BibitemShut {NoStop}%
\bibitem [{\citenamefont {B{\"o}hm}\ and\ \citenamefont
  {Szlach{\'o}nyi}(1996)}]{bohm1996coassociative}%
  \BibitemOpen
  \bibfield  {author} {\bibinfo {author} {\bibfnamefont {G.}~\bibnamefont
  {B{\"o}hm}}\ and\ \bibinfo {author} {\bibfnamefont {K.}~\bibnamefont
  {Szlach{\'o}nyi}},\ }\bibfield  {title} {\enquote {\bibinfo {title} {A
  coassociative ${C}^*$-quantum group with nonintegral dimensions},}\ }\href
  {https://link.springer.com/article/10.1007/BF01815526} {\bibfield  {journal}
  {\bibinfo  {journal} {Letters in Mathematical Physics}\ }\textbf {\bibinfo
  {volume} {38}},\ \bibinfo {pages} {437} (\bibinfo {year} {1996})},\ \Eprint
  {http://arxiv.org/abs/q-alg/9509008} {arXiv:q-alg/9509008 [math.QA]}
  \BibitemShut {NoStop}%
\bibitem [{\citenamefont {Szlach{\'a}nyi}(2001)}]{szlachanyi2000finite}%
  \BibitemOpen
  \bibfield  {author} {\bibinfo {author} {\bibfnamefont {K.}~\bibnamefont
  {Szlach{\'a}nyi}},\ }\bibfield  {title} {\enquote {\bibinfo {title} {Finite
  quantum groupoids and inclusions of finite type},}\ }in\ \href@noop {} {\emph
  {\bibinfo {booktitle} {Mathematical physics in mathematics and physics:
  quantum and operator algebraic aspects}}},\ \bibinfo {series} {Fields
  Institute Communications}, Vol.~\bibinfo {volume} {30}\ (\bibinfo
  {publisher} {American Mathematical Soc.},\ \bibinfo {year} {2001})\ p.\
  \bibinfo {pages} {393–407},\ \Eprint {http://arxiv.org/abs/math/0011036}
  {arXiv:math/0011036 [math.QA]} \BibitemShut {NoStop}%
\bibitem [{\citenamefont {Ostrik}(2003)}]{ostrik2003module}%
  \BibitemOpen
  \bibfield  {author} {\bibinfo {author} {\bibfnamefont {V.}~\bibnamefont
  {Ostrik}},\ }\bibfield  {title} {\enquote {\bibinfo {title} {Module
  categories, weak {H}opf algebras and modular invariants},}\ }\href
  {https://link.springer.com/content/pdf/10.1007%2Fs00031-003-0515-6.pdf}
  {\bibfield  {journal} {\bibinfo  {journal} {Transformation Groups}\ }\textbf
  {\bibinfo {volume} {8}},\ \bibinfo {pages} {177} (\bibinfo {year} {2003})},\
  \Eprint {http://arxiv.org/abs/math/0111139} {arXiv:math/0111139 [math.QA]}
  \BibitemShut {NoStop}%
\bibitem [{\citenamefont {Kitaev}\ and\ \citenamefont
  {Kong}(2012)}]{Kitaev2012boundary}%
  \BibitemOpen
  \bibfield  {author} {\bibinfo {author} {\bibfnamefont {A.}~\bibnamefont
  {Kitaev}}\ and\ \bibinfo {author} {\bibfnamefont {L.}~\bibnamefont {Kong}},\
  }\bibfield  {title} {\enquote {\bibinfo {title} {Models for gapped boundaries
  and domain walls},}\ }\href {\doibase 10.1007/s00220-012-1500-5} {\bibfield
  {journal} {\bibinfo  {journal} {Communications in Mathematical Physics}\
  }\textbf {\bibinfo {volume} {313}},\ \bibinfo {pages} {351} (\bibinfo {year}
  {2012})},\ \Eprint {http://arxiv.org/abs/1104.5047} {arXiv:1104.5047
  [cond-mat.str-el]} \BibitemShut {NoStop}%
\bibitem [{\citenamefont {Bridgeman}\ \emph {et~al.}(2023)\citenamefont
  {Bridgeman}, \citenamefont {Lootens},\ and\ \citenamefont
  {Verstraete}}]{bridgeman2023invertible}%
  \BibitemOpen
  \bibfield  {author} {\bibinfo {author} {\bibfnamefont {J.~C.}\ \bibnamefont
  {Bridgeman}}, \bibinfo {author} {\bibfnamefont {L.}~\bibnamefont {Lootens}},
  \ and\ \bibinfo {author} {\bibfnamefont {F.}~\bibnamefont {Verstraete}},\
  }\bibfield  {title} {\enquote {\bibinfo {title} {Invertible bimodule
  categories and generalized {S}chur orthogonality},}\ }\href
  {https://link.springer.com/article/10.1007/s00220-023-04781-y} {\bibfield
  {journal} {\bibinfo  {journal} {Communications in Mathematical Physics}\
  }\textbf {\bibinfo {volume} {402}},\ \bibinfo {pages} {2691} (\bibinfo {year}
  {2023})},\ \Eprint {http://arxiv.org/abs/2211.01947} {arXiv:2211.01947
  [math.QA]} \BibitemShut {NoStop}%
\bibitem [{\citenamefont {Bai}\ and\ \citenamefont
  {Zhang}(2025)}]{bai2025weakhopf}%
  \BibitemOpen
  \bibfield  {author} {\bibinfo {author} {\bibfnamefont {A.}~\bibnamefont
  {Bai}}\ and\ \bibinfo {author} {\bibfnamefont {Z.-H.}\ \bibnamefont
  {Zhang}},\ }\href {https://arxiv.org/abs/2503.06731} {\enquote {\bibinfo
  {title} {On the representation categories of weak {H}opf algebras arising
  from {L}evin-{W}en models},}\ } (\bibinfo {year} {2025}),\ \Eprint
  {http://arxiv.org/abs/2503.06731} {arXiv:2503.06731 [math.QA]} \BibitemShut
  {NoStop}%
\bibitem [{\citenamefont {Drinfel'd}(1988)}]{drinfel1988quantum}%
  \BibitemOpen
  \bibfield  {author} {\bibinfo {author} {\bibfnamefont {V.~G.}\ \bibnamefont
  {Drinfel'd}},\ }\bibfield  {title} {\enquote {\bibinfo {title} {Quantum
  groups},}\ }\href {https://link.springer.com/article/10.1007/BF01247086}
  {\bibfield  {journal} {\bibinfo  {journal} {Journal of Soviet mathematics}\
  }\textbf {\bibinfo {volume} {41}},\ \bibinfo {pages} {898} (\bibinfo {year}
  {1988})}\BibitemShut {NoStop}%
\bibitem [{\citenamefont {Majid}(1990)}]{majid1990physics}%
  \BibitemOpen
  \bibfield  {author} {\bibinfo {author} {\bibfnamefont {S.}~\bibnamefont
  {Majid}},\ }\bibfield  {title} {\enquote {\bibinfo {title} {Physics for
  algebraists: Non-commutative and non-cocommutative {H}opf algebras by a
  bicrossproduct construction},}\ }\href
  {https://www.sciencedirect.com/science/article/pii/002186939090099A}
  {\bibfield  {journal} {\bibinfo  {journal} {Journal of Algebra}\ }\textbf
  {\bibinfo {volume} {130}},\ \bibinfo {pages} {17} (\bibinfo {year}
  {1990})}\BibitemShut {NoStop}%
\bibitem [{\citenamefont {Majid}(1994)}]{majid1994some}%
  \BibitemOpen
  \bibfield  {author} {\bibinfo {author} {\bibfnamefont {S.}~\bibnamefont
  {Majid}},\ }\bibfield  {title} {\enquote {\bibinfo {title} {Some remarks on
  the quantum double},}\ }\href
  {https://link.springer.com/article/10.1007/BF01690458} {\bibfield  {journal}
  {\bibinfo  {journal} {Czechoslovak Journal of Physics}\ }\textbf {\bibinfo
  {volume} {44}},\ \bibinfo {pages} {1059} (\bibinfo {year} {1994})},\ \Eprint
  {http://arxiv.org/abs/hep-th/9409056} {arXiv:hep-th/9409056 [hep-th]}
  \BibitemShut {NoStop}%
\bibitem [{\citenamefont {Nikshych}\ and\ \citenamefont
  {Vainerman}(2002)}]{nikshych2002finite}%
  \BibitemOpen
  \bibfield  {author} {\bibinfo {author} {\bibfnamefont {D.}~\bibnamefont
  {Nikshych}}\ and\ \bibinfo {author} {\bibfnamefont {L.}~\bibnamefont
  {Vainerman}},\ }\bibfield  {title} {\enquote {\bibinfo {title} {Finite
  quantum groupoids and their applications},}\ }\href@noop {} {\bibfield
  {journal} {\bibinfo  {journal} {New directions in Hopf algebras}\ }\textbf
  {\bibinfo {volume} {43}},\ \bibinfo {pages} {211} (\bibinfo {year} {2002})},\
  \Eprint {http://arxiv.org/abs/math/0006057} {arXiv:math/0006057 [math.QA]}
  \BibitemShut {NoStop}%
\bibitem [{\citenamefont {Kassel}(1995)}]{kassel2012quantum}%
  \BibitemOpen
  \bibfield  {author} {\bibinfo {author} {\bibfnamefont {C.}~\bibnamefont
  {Kassel}},\ }\href {https://link.springer.com/book/10.1007/978-1-4612-0783-2}
  {\emph {\bibinfo {title} {Quantum groups}}},\ \bibinfo {series} {Graduate
  Texts in Mathematics}, Vol.\ \bibinfo {volume} {155}\ (\bibinfo  {publisher}
  {Springer-Verlag, New York},\ \bibinfo {year} {1995})\ pp.\ \bibinfo {pages}
  {xii+531}\BibitemShut {NoStop}%
\bibitem [{\citenamefont {Bravyi}\ and\ \citenamefont
  {Kitaev}(1998)}]{bravyi1998quantum}%
  \BibitemOpen
  \bibfield  {author} {\bibinfo {author} {\bibfnamefont {S.~B.}\ \bibnamefont
  {Bravyi}}\ and\ \bibinfo {author} {\bibfnamefont {A.~Y.}\ \bibnamefont
  {Kitaev}},\ }\bibfield  {title} {\enquote {\bibinfo {title} {Quantum codes on
  a lattice with boundary},}\ }\href@noop {} {\  (\bibinfo {year} {1998})},\
  \Eprint {http://arxiv.org/abs/quant-ph/9811052} {arXiv:quant-ph/9811052
  [quant-ph]} \BibitemShut {NoStop}%
\bibitem [{\citenamefont {Chang}(2014)}]{chang2014kitaev}%
  \BibitemOpen
  \bibfield  {author} {\bibinfo {author} {\bibfnamefont {L.}~\bibnamefont
  {Chang}},\ }\bibfield  {title} {\enquote {\bibinfo {title} {Kitaev models
  based on unitary quantum groupoids},}\ }\href
  {https://aip.scitation.org/doi/abs/10.1063/1.4869326} {\bibfield  {journal}
  {\bibinfo  {journal} {Journal of Mathematical Physics}\ }\textbf {\bibinfo
  {volume} {55}},\ \bibinfo {pages} {041703} (\bibinfo {year} {2014})},\
  \Eprint {http://arxiv.org/abs/1309.4181} {arXiv:1309.4181 [math.QA]}
  \BibitemShut {NoStop}%
\end{thebibliography}%

\end{document}